\documentclass[screen, 10pt]{acmart}
\usepackage{algorithm}
\usepackage{algorithmic}
\usepackage{balance}
\usepackage{blindtext}
\usepackage{amsmath,amssymb,amsfonts}
\usepackage{algorithmic}
\usepackage{graphicx}
\usepackage{xspace}
\usepackage{tabularx}
\usepackage{array}
\usepackage{booktabs}
\newcommand{\midsepremove}{\aboverulesep=0mm \belowrulesep=0mm}
\midsepremove
\newcommand{\midsepdefault}{\aboverulesep=0.6mm \belowrulesep=0.9mm}
\midsepdefault
\usepackage{xcolor}
\usepackage{colortbl}
\usepackage{multirow}
\usepackage{algorithm}
\usepackage{algorithmic}

\newcommand*{\eg}{\textit{e.g.}\@\xspace}
\newcommand*{\ie}{\textit{i.e.}\@\xspace}
\makeatletter
\newcommand*{\etc}{%
    \@ifnextchar{.}%
        {etc}%
        {etc.\@\xspace}%
}
\newcommand*{\etal}{%
    \@ifnextchar{.}%
        {et al}%
        {et al.\@\xspace}%
}
\makeatother

\newboolean{showcomments}
\setboolean{showcomments}{true} 
\ifthenelse{\boolean{showcomments}}
{ \newcommand{\mynote}[3]{
    \fbox{\bfseries\sffamily\scriptsize#1}
    {\small$\blacktriangleright$\textsf{\emph{\color{#3}{#2}}}$\blacktriangleleft$}}}
{ \newcommand{\mynote}[3]{}}
\newcommand{\shrink}[1]{}

\graphicspath{{./figs/}}

\newcommand{\topone} {{top-1}\xspace}
\newcommand{\topfive} {{top-5}\xspace}
\newcommand{\DNN} {\texttt{DNN}\xspace}
\newcommand{\DNNs} {\texttt{DNNs}\xspace}
\newcommand{\CNN} {\texttt{CNN}\xspace}
\newcommand{\RNN} {\texttt{RNN}\xspace}

\newcommand{\CNNs} {\texttt{CNNs}\xspace}
\newcommand{\NN} {\texttt{KNN}\xspace}

\newcommand{\SVM} {\texttt{SVM}\xspace}
\newcommand{\DT} {\texttt{DT}\xspace}

\newcommand{\mn}[1]{\texttt{#1}}
\newcommand{\premodel} {\texttt{premodel}\xspace}

\newcommand{\cparagraph}[1]{\vspace{1mm}\noindent \textbf{#1}}

\definecolor{Gray}{gray}{0.95}

\setcopyright{acmcopyright}
\acmDOI{10.475/123_4}
\acmISBN{123-4567-24-567/08/06}
\acmConference[LCTES '18]{}{June 2018}{Pennsylvania, USA}

\emergencystretch=\maxdimen
\begin{document}

\setcopyright{iw3c2w3g}

\title{Adaptive Selection of Deep Learning Models on Embedded Systems}

\author{Ben Taylor, Vicent Sanz Marco, Willy Wolff, Yehia Elkhatib, Zheng Wang}
\affiliation{\institution{MetaLab, School of Computing and Communications, Lancaster University, United Kingdom}}
\email{{b.d.taylor,
v.sanzmarco, w.wolff, y.elkhatib, z.wang}@lancaster.ac.uk}

%
%
%
%

\thanks{Ben Taylor and Vicent Sanz Marco provided an equal contribution.}

\renewcommand{\shortauthors}{B. Taylor et al.}

\begin{abstract}
The recent ground-breaking advances in deep learning networks (\DNNs) make them attractive for embedded systems. However, it can take a
long time for \DNNs to make an inference on resource-limited embedded devices. Offloading the computation into the cloud is often
infeasible due to privacy concerns, high latency, or the lack of connectivity. As such, there is a critical need to find a way to
effectively execute the \DNN models locally on the devices.

This paper presents an adaptive scheme to determine which \DNN model to use for a given input, by considering the desired accuracy and
inference time. Our approach employs machine learning to develop a predictive model to quickly select a pre-trained \DNN to use for a
given input and the optimization constraint. We achieve this by first training off-line a predictive model, and then use the learnt model
to select a \DNN model to use for new, unseen inputs. We apply our approach to the image classification task and evaluate it on a Jetson
TX2 embedded deep learning platform using the ImageNet ILSVRC 2012 validation dataset. We consider a range of influential \DNN models.
Experimental results show that our approach achieves a 7.52\% improvement in inference accuracy, and a 1.8x reduction in inference time
over the most-capable single \DNN model.
\end{abstract}


 \maketitle

\vspace{-1mm}
\section{Introduction}
Recent advances in deep learning have brought a step change in the abilities of machines in solving complex problems like object
recognition~\cite{donahue14,he2016deep}, facial recognition~\cite{parkhi2015deep,sun2014deep}, speech processing~\cite{pmlrv48amodei16},
and machine translation~\cite{bahdanau2014neural}. Although many of these tasks are important on mobile and embedded devices, especially
for sensing and mission critical applications such as health care and video surveillance, existing deep learning solutions often require a
large amount of computational resources to run. Running these models on embedded devices can lead to long runtimes and the consumption of
abundant amounts of resources, including CPU, memory, and power, even for simple tasks~\cite{CanzianiPC16}. Without a solution,
 the hoped-for advances on embedded sensing will not arrive.

A common approach for accelerating \DNN models on embedded devices is to compress the model to reduce its resource and computational
requirements~\cite{han2015learning,han2016eie,howard2017mobilenets,Georgiev:2017:LMA:3139486.3131895}, but this comes at the cost of a loss
in precision. Other approaches involve offloading some, or all, computation to a cloud
server~\cite{Kang2017neurosurgeon,teerapittayanon2017distributed}. This, however, is not always possible due to constraints on privacy,
when sending sensitive data over the network is prohibitive; and latency, where a fast, reliable network connection is not always
guaranteed.

\begin{figure*}[t!]
\def\arraystretch{0.8}
    \centering
    \begin{tabularx}{1\textwidth} {>{\centering\arraybackslash}m{1.4in}>{\centering\arraybackslash}m{1.4in}>{\centering\arraybackslash}m{1.4in}>{\centering\arraybackslash}m{2.5in}}
        \includegraphics[width=0.2\textwidth]{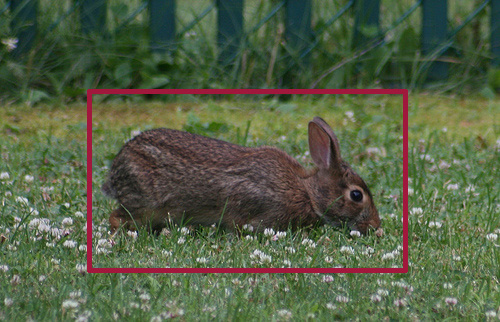} &
        \includegraphics[width=0.2\textwidth]{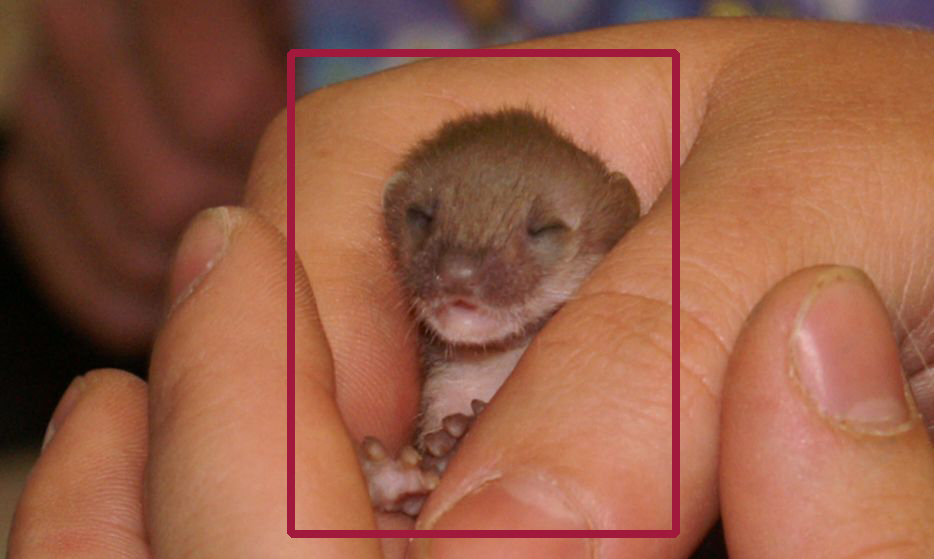} &
        \includegraphics[width=0.2\textwidth]{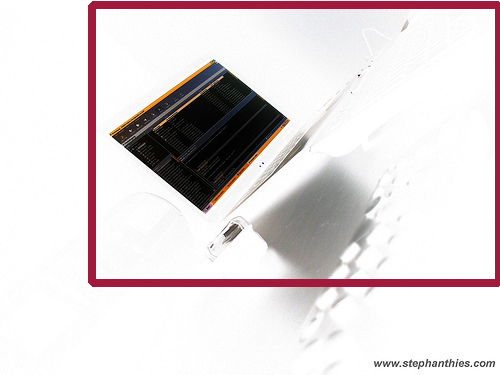} &
        \includegraphics[width=0.29\textwidth]{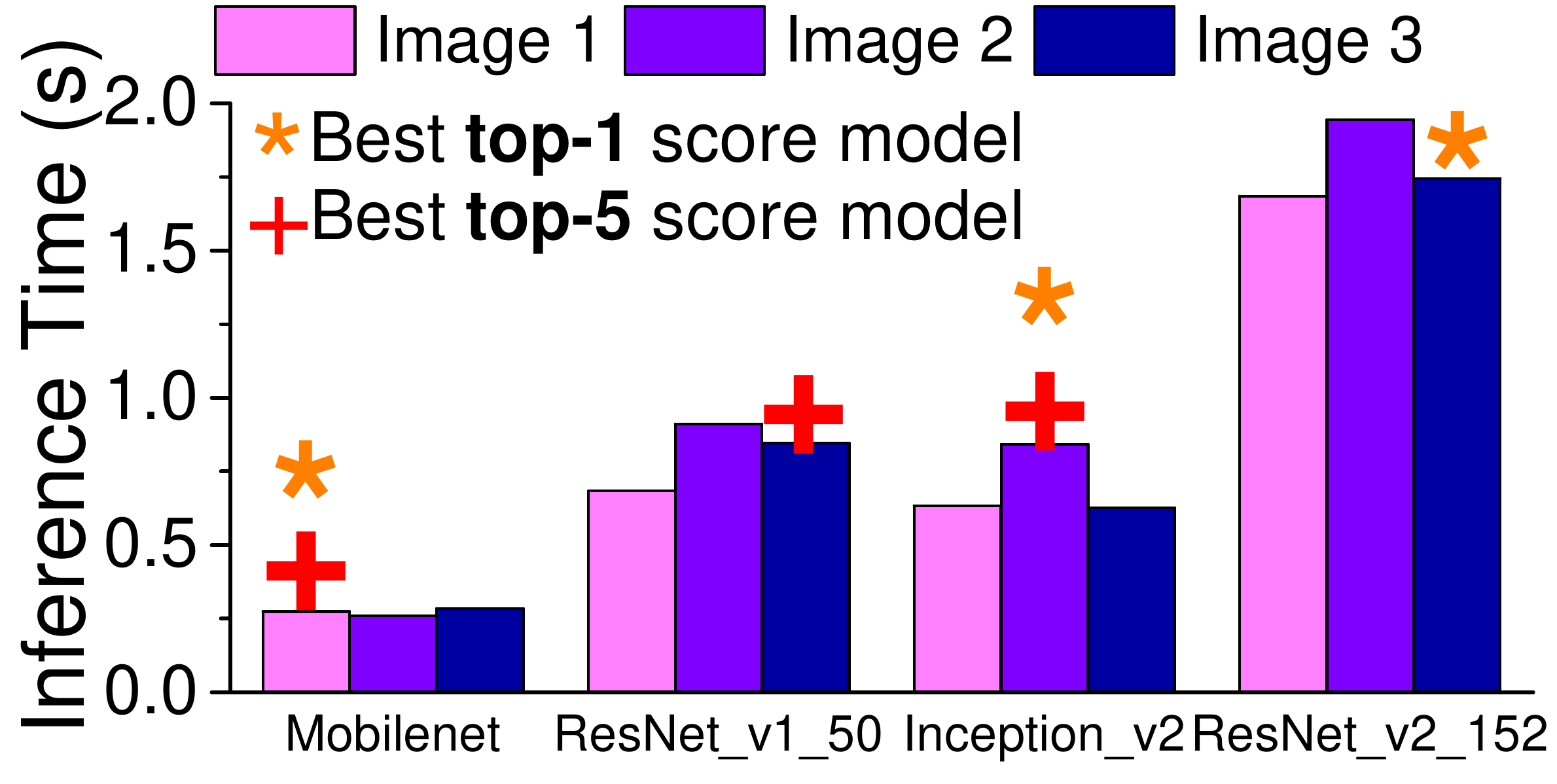} \\
        \vspace{-2mm}
        {\centering \scriptsize (a) Image 1} &
        {\centering \scriptsize (b) Image 2} &
        {\centering \scriptsize (c) Image 3} &
        {\centering \scriptsize (d) Inference time}
    \end{tabularx}
    \vspace{-4mm}
    \caption{The inference time (d) of four CNN-based image recognition models when processing images (a) - (c).
    The target object is highlighted on each image.
    This example (combined with Table~\ref{tbl:motivation_model_accuracy}) shows that the best model to use (\ie the fastest model that gives the accurate output)
    depends on the success criterion and the input.}
    \label{fig:motivation_example}
\end{figure*}

This paper seeks to offer an alternative to enable efficient deep \emph{inference}\footnote{Inference in this work means applying a
pre-trained model on an input to obtain the corresponding output. This is different from statistical inference.} on embedded devices.  Our
goal is to design an adaptive scheme to determine, \emph{at runtime}, which of the available \DNN models is the best fit for the input and
the precision requirement. This is motivated by the observation that the optimum model\footnote{In this work, the optimum model is the one
that gives the correct output with the fastest inference time.}  for inference depends on the input data and the precision requirement. For
example, if the input image is taken under good lighting conditions with a simple background, a simple but fast model would be
sufficient for object identification -- otherwise, a more sophisticated but slower model will have to be employed; in a
similar vein, if we want to detect objects with a high confidence, an advanced model should be used -- otherwise, a simple model
would be good enough. Given that \DNN models are becoming increasingly diverse -- together with the evolving application workload and user
requirements, the right strategy for model selection is likely to change over time. This ever-evolving nature makes automatic heuristic
design highly attractive because the heuristic can be easily updated to adapt to the changing application context.

This paper presents a novel runtime approach for \DNN model selection on embedded devices, aiming to minimize the inference time while
meeting the user requirement. We achieve this by employing machine learning to \emph{automatically} construct predictors to select at
runtime the optimum model to use. Our predictor is first trained \emph{off-line}. Then, using a set of automatically tuned features of the
\DNN model input, the predictor determines the optimum \DNN model for a \emph{new}, \emph{unseen} input, by taking into consideration the
precision constraint and the characteristics of the input. We show that our approach can automatically derive high-quality heuristics for
different precision requirements. The learned strategy can effectively leverage the prediction capability and runtime overhead of candidate
\DNN models, leading to an overall better accuracy when compared with the most capable \DNN model, but with significantly less runtime
overhead. Using our approach, one can also first apply model compression techniques to generate \DNN models of different capabilities and
inference time, and then choose a model to use at runtime. This is a new way for optimizing deep inference on embedded devices.

We apply our approach to the image classification domain, an area where deep learning has made impressive breakthroughs by using
high-performance systems and where a rich set of pre-trained models are available. We evaluate our approach on the NVIDIA Jetson TX2
embedded deep learning platform and consider a wide range of influential \DNN models. Our experiments are performed using the 50K images
from the ImageNet ILSVRC 2012 validation dataset. To show the automatic portability of our approach across precision requirements, we have
evaluated it on two different evaluation criteria used by the ImageNet contest. Our approach is able to correctly choose the optimum model
to use for 95.6\% of the test cases. Overall, it improves the
inference accuracy by 7.52\% over the most-capable single model but with 1.8x less inference time.

The paper makes the following contributions:
\vspace{-1mm}
\begin{itemize}
\item We present a novel machine learning based approach to automatically learn how to select \DNN models based on the input and
    precision requirement (Section~\ref{sec:approach});

\item Our work is the first to leverage multiple \DNN models to improve the prediction accuracy and reduce inference time on embedded
    systems (Section~\ref{sec:results}). Our approach allows developers to easily re-target the approach for new \DNN models and user
    requirements;

\item Our system has little training overhead  as it does not require any modification to pre-trained \DNN models.

\end{itemize}

\section{Motivation and Overview}
%
\subsection{Motivation}  \label{sec:motivation}

\begin{table}[t!]
	\scriptsize
	\centering
	\caption{List of models that give the correct prediction per image under the \topfive and
    the \topone scores.}
	\vspace{-2mm}
	\midsepremove
	\begin{tabularx}{0.48\textwidth}{p{1.2cm}*{3}{X}}
		\toprule
             & \topfive score & \topone score\\
            \midrule
            Image 1 & \mn{\textbf{MobileNet\_v1\_025}}, \mn{ResNet\_v1\_50}, \mn{Inception\_v2}, \mn{ResNet\_v2\_152} & \mn{\textbf{MobileNet\_v1\_025}}, \mn{ResNet\_v1\_50}, \mn{Inception\_v2}, \mn{ResNet\_v2\_152} \\
            \rowcolor{Gray} Image 2 & \mn{\textbf{Inception\_v2}}, \mn{ResNet\_v1\_50},  \mn{ResNet\_v2\_152} & \mn{\textbf{Inception\_v2}}, \mn{ResNet\_v2\_152}\\
            \rowcolor{white} Image 3 & \mn{\textbf{ResNet\_v1\_50}}, \mn{ResNet\_v2\_152} & \mn{\textbf{ResNet\_v2\_152}} \\
            \bottomrule
	\end{tabularx}
	\label{tbl:motivation_model_accuracy}
\vspace{-2mm}
\end{table}

As a motivating example, consider performing object recognition on a NVIDIA Jetson TX2 platform.

\cparagraph{Setup.} In this experiment, we compare the performance of three influential Convolutional Neural Network (\CNN) architectures:
\mn{Inception}~\cite{ioffe2015batch}, \mn{ResNet}~\cite{he2016identity}, and \mn{MobileNet}~\cite{howard2017mobilenets}\footnote{
Each model architecture follows its own naming convention. \mn{MobileNet\_v$i$\_$j$},
 where $i$ is the version number, and $j$ is a width multiplier out of 100, with 100 being the full uncompressed model.
\mn{ResNet\_v$i$\_$j$}, where $i$ is the version number, and $j$ is the number of layers in the model.
\mn{Inception\_v$i$}, where $i$ is the version number.}.
Specifically, we
used the following models: \mn{MobileNet\_v1\_025},
the \mn{MobileNet} architecture with a width multiplier of 0.25;
\mn{ResNet\_v1\_50},
the first version of \mn{ResNet} with 50 layers;
\mn{Inception\_v2},
the second version of \mn{Inception};
and \mn{ResNet\_v2\_152},
the second version of \mn{ResNet} with 152 layers.
All these models are built upon TensorFlow~\cite{tensorflow} and  have been pre-trained by
independent researchers using the ImageNet ILSVRC 2012 \emph{training dataset}~\cite{ILSVRC15}. We use the GPU for inference.

\cparagraph{Evaluation Criteria.} Each model takes an image as input and returns a list of label confidence values as output. Each value
indicates the confidence that a particular object is in the image. The resulting list of object values are sorted in descending order
regarding their prediction confidence, so that the label with the highest confidence appears at the top of the list. In this example, the
accuracy of a model is evaluated using the \topone and the \topfive scores defined by the ImageNet Challenge. Specifically, for the \topone
score, we check if the top output label matches the ground truth label of the primary object; and for the \topfive score, we check if the
ground truth label of the primary object is in the top 5 of the output labels for each given model.

\cparagraph{Results.} Figure~\ref{fig:motivation_example}d shows the inference time per model using three images from the ImageNet ILSVRC
\emph{validation dataset}. Recognizing the main object (a cottontail rabbit) from the image shown in Figure~\ref{fig:motivation_example}a
is a straightforward task. We can see from Figure~\ref{tbl:motivation_model_accuracy} that all models give the correct answer under the
\topfive and \topone score criterion. For this image, \mn{MobileNet\_v1\_025} is the best model to use under the \topfive score, because it
has the fastest inference time -- 6.13x faster than \mn{ResNet\_v2\_152}. Clearly, for this image, \mn{MobileNet\_v1\_025} is good enough,
and there is no need to use a more advanced (and expensive) model for inference. If we consider a slightly more complex object
recognition task shown in Figure~\ref{fig:motivation_example}b, we can see that \mn{MobileNet\_v1\_025} is unable to give a correct answer
regardless of our success criterion. In this case \mn{Inception\_v2} should be used, although this is 3.24x slower than
\mn{MobileNet\_v1\_025}. Finally, consider the image shown in Figure~\ref{fig:motivation_example}c, intuitively it can be seen that
this is a more difficult image recognition task, the main object is a similar color to the background. In this case the optimal model
changes depending on our success criterion. \mn{ResNet\_v1\_50} is the best model to use under \topfive scoring, completing
inference 2.06x faster than \mn{ResNet\_v2\_152}. However, if we use \topone for scoring we must use \mn{ResNet\_v2\_152}  to obtain the
correct label, despite it being the most expensive model. Inference time for this image is 2.98x and 6.14x slower than
\mn{MobileNet\_v1\_025} for \topfive and \topone scoring respectively.

\cparagraph{Lessons Learned.} This example shows that the best model depends on the input and the evaluation criterion. Hence, determining
which model to use is non-trivial. What we need is a technique that can automatically choose the most efficient model to use for any given
input. In the next section, we describe our adaptive approach that solves this task.

\subsection{Overview of Our Approach}
\begin{figure}[t!]
  \centering
  \includegraphics[width=0.7\textwidth]{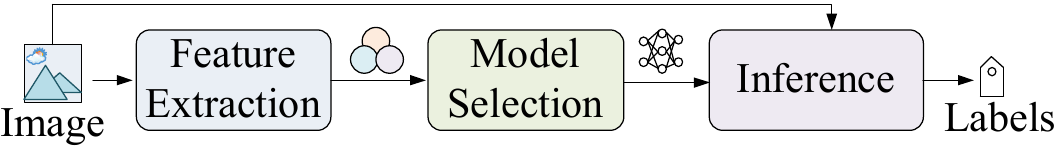}\\
  \caption{Overview of our approach} \label{fig:overview}
\end{figure}

Figure~\ref{fig:overview} depicts the overall work flow of our approach. While our approach is generally applicable, to have a concrete,
measurable target, we apply it to image classification. At the core of our approach is a predictive model (termed \premodel) that takes a
\emph{new, unseen} image to predict which of a set of pre-trained image classification models to use for the given input. This decision may
vary depending on the scoring method used at the time, \eg either \topone or \topfive, and we show that our approach can adapt to
different metrics.

The prediction of our \premodel is based on a set of quantifiable properties -- or \emph{features} such as the number of edges and
brightness -- of the input image. Once a model is chosen, the input image is passed to the selected model, which then attempts to classify
the image. Finally, the classification data of the selected model is returned as outputs. Use of our \premodel will work in exactly the
same way as any single model, the difference being we are able to choose the best model to use dynamically.


\section{Our Approach \label{sec:approach}}

\begin{figure}[t!]
  \centering
  \includegraphics[width=0.7\textwidth]{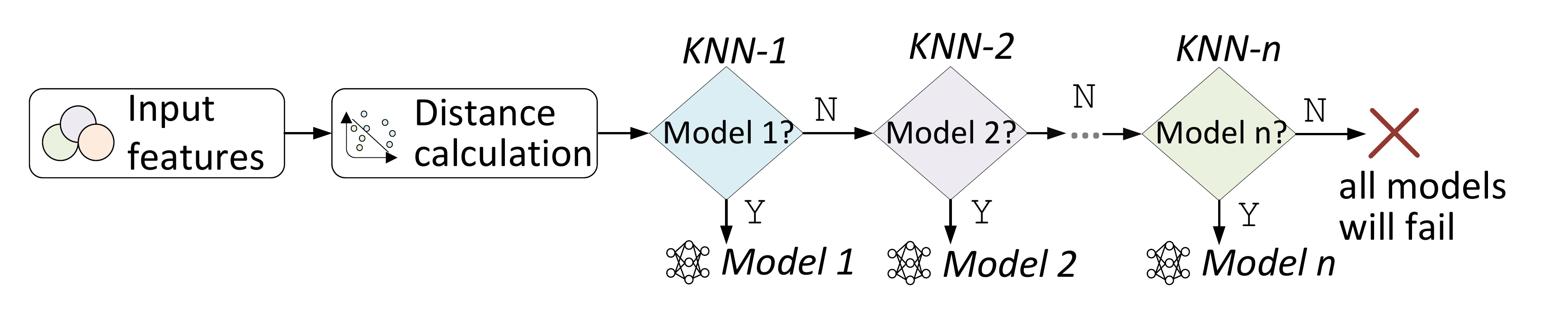}\\
  \caption{Our \premodel, made up of a series of \NN models. Each model predicts whether to use an image classifier or not, our selection
  process for including image classifiers is described in Section~\ref{sec:classifier_selection}.}
  \label{fig:premodel_design}
\end{figure}

Our \premodel is made up of multiple k-Nearest Neighbour (\NN) classification models arranged in sequence, shown in
Figure~\ref{fig:premodel_design}\footnote{In Section~\ref{sec:alt_premodels}, we evaluate a number of different machine learning
techniques, including Decision Trees, Support Vector Machines, and \CNNs.}. As input our model takes an image, from which it will extract
features and make a prediction, outputting a label referring to which image classification model to use.

\subsection{Model Description} \label{sec:model_desc}

There are two main requirements to consider when developing an inferencing model selection strategy on an embedded device: (i) fast
execution time, and (ii) high accuracy. A \premodel which takes much longer than any single model would outweigh its benefits.
High accuracy leads to a reduced overall cost, that is, the average inference time for any single image.

We chose to implement a series of simple \NN models, where each model predicts whether to use a single image
classifier or not. We chose \NN as it has a quick prediction time (<1ms) and achieves a high accuracy for our problem. Finally, we
chose a set of features to represent each image; the selection process is described in more detail in
Section~\ref{sec:features}.

Figure~\ref{fig:premodel_design} gives an overview of our \premodel architecture. For each \DNN model we wish to include in our \premodel,
we use a separate \NN model. As our \NN models are going to contain much of the same data we begin our \premodel by calculating our K
closest neighbours. Taking note of which record of training data each of the neighbours corresponds to, we are able to avoid recalculating
the distance measurements; we simply change the labels of these data-points. \textit{KNN-1} is the first \NN model in our \premodel,
through which all input to the \premodel will pass. \textit{KNN-1} predicts whether the input image should use \textit{Model-1}
to classify it or not, depending on the scoring criterion the \premodel has been trained for. If \textit{KNN-1} predicts that
\textit{Model-1} should be used, then the \premodel returns this label, otherwise the features are passed on to the next \NN, \ie
\textit{KNN-2}. This process repeats until the image reaches \textit{KNN-n}, the final \NN model in our \premodel. In the event that
\textit{KNN-n} predicts that we should not use \textit{Model-n} to classify the image, the next step will depend on the
user's declared preference: (i) use a pre-specified model, to receive some output to work with; or (ii) do not perform
inference and inform the user of the failure.



\begin{figure}[t!]
  \centering
  \includegraphics[width=0.7\textwidth]{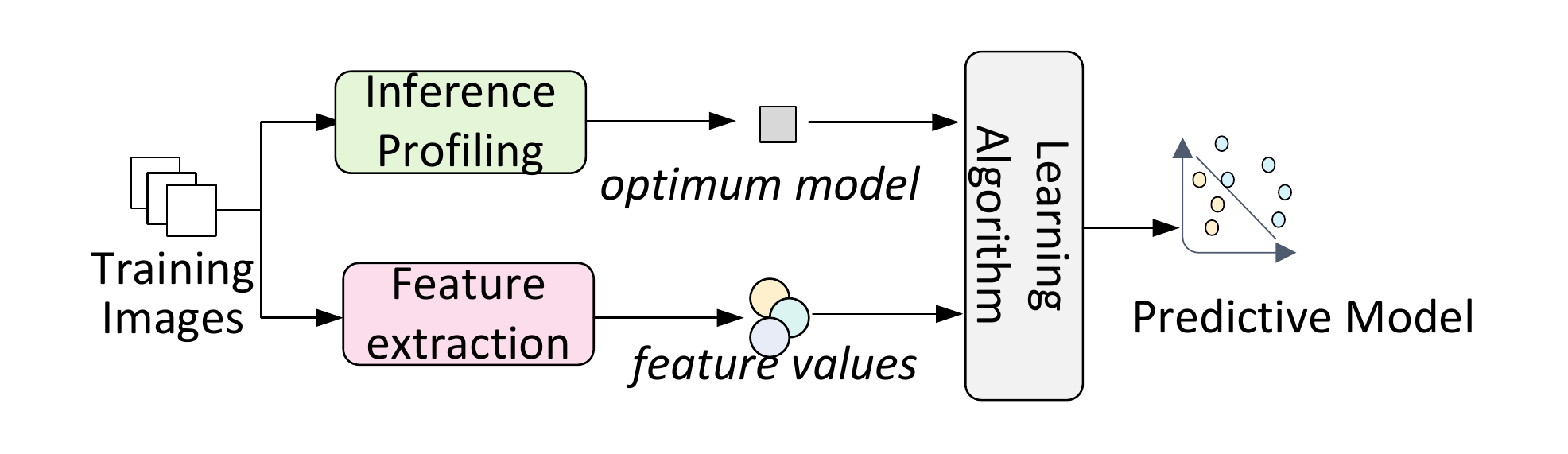}\\
  \vspace{-2mm}
  \caption{The training process. We use the same procedure to train each individual model within the \premodel for each evaluation criterion.}\label{fig:training}
\end{figure}

\begin{table*}[t!]
\def\arraystretch{0.82}
	\scriptsize
	\centering
	\caption{All features considered in this work.}
	\vspace{-2mm}
	\midsepremove
	\begin{tabular}{llll}
		\toprule
							\textbf{Feature}                    & \textbf{Description}             &         \textbf{Feature}                    & \textbf{Description}  \\
		\midrule
		\rowcolor{Gray}		\textit{n\_keypoints}               & \# of keypoints                                         &
							\textit{avg\_brightness}            & Average brightness                                      \\
		                    \textit{brightness\_rms}            & Root mean square of brightness                          &
							\textit{avg\_perceived\_brightness} & Average of perceived brightness                         \\
		\rowcolor{Gray}		\textit{perceived\_brightness\_rms} & Root mean square of perceived  brightness               &
							\textit{contrast}                   & The level of contrast                                   \\
	                   		\textit{edge\_length\{1-7\}}        & A 7-bin histogram of edge lengths                       &
							\textit{edge\_angle\{1-7\}}         & A 7-bin histogram of edge angles                        \\
		\rowcolor{Gray}		\textit{area\_by\_perim}            & Area / perimeter of the main object                     &
							\textit{aspect\_ratio}              & The aspect ratio of the main object                     \\
		              		\textit{hue\{1-7\}}                 & A 7-bin histogram of the different hues & &             \\
		\bottomrule
	\end{tabular}
	\label{tbl:all_features}
\end{table*}

\begin{algorithm}[t!]
	\caption{Inference Model Selection Process}
	\label{alg:classifier_selection}
    \small
	\begin{algorithmic}
	\STATE $Model\_1\_DNN = most\_optimum\_DNN(data)$
	\STATE $curr\_DNNs.add(Model\_1\_DNN)$
	\STATE $curr\_acc = get\_acc(curr\_DNNs)$
	\STATE \textit{acc\_diff = 100}
	\WHILE{$\textit{acc\_diff} > \theta$}
		\STATE \textit{failed\_cases = get\_fail\_cases(curr\_DNNs)}
		\STATE \textit{next\_DNN = most\_acc\_DNN(failed\_cases)}
		\STATE $curr\_DNNs.add(next\_DNN)$
		\STATE $new\_acc = get\_acc(curr\_DNNs)$
		\STATE \textit{acc\_diff = new\_acc - curr\_acc}
		\STATE $curr\_acc = new\_acc$
	\ENDWHILE
	\end{algorithmic}
\end{algorithm}

\subsection{Inference Model Selection} \label{sec:classifier_selection}

\begin{figure}[t!]
	\centering
	\includegraphics[width=0.7\textwidth]{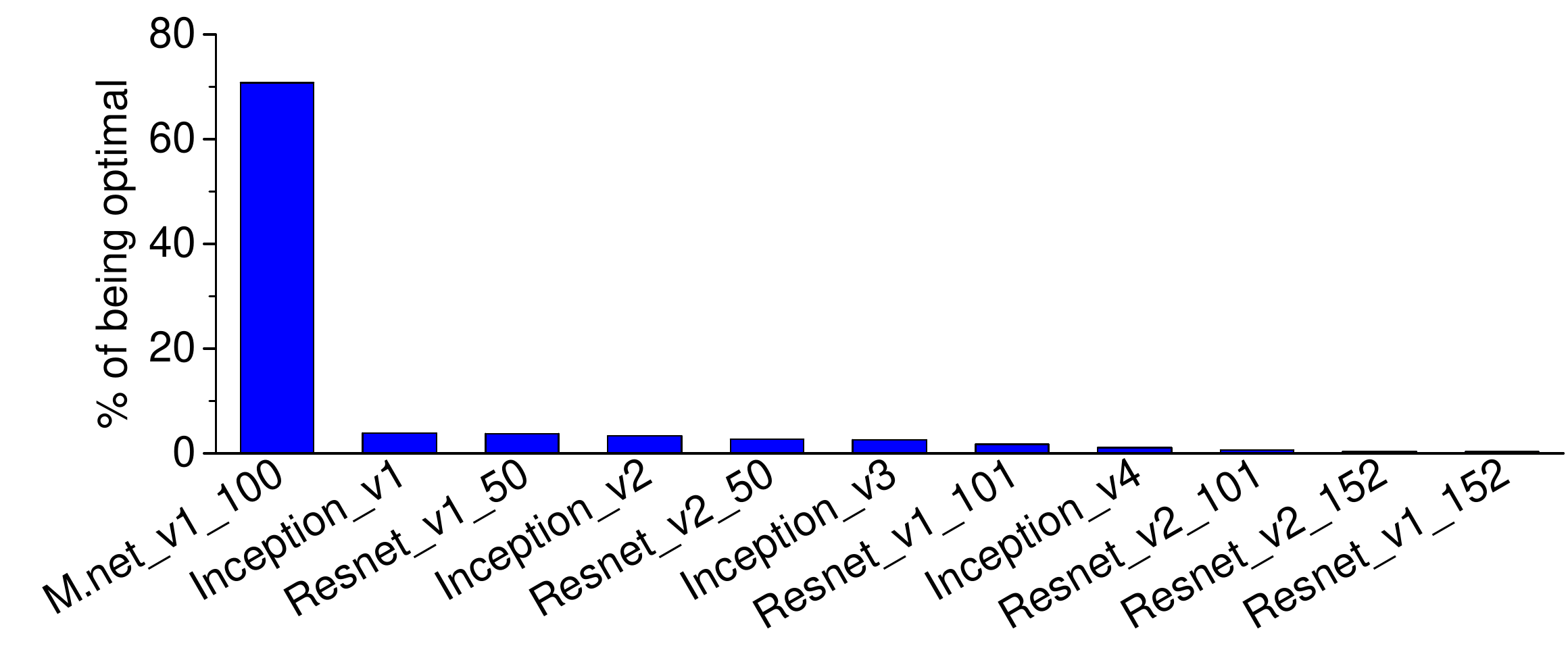}\\
    \vspace{-2mm}
	\caption{How often a \CNN model is considered to be optimal under the \topone score on the training dataset.}
	\label{fig:optimum_models}
\end{figure}

In Algorithm~\ref{alg:classifier_selection} we describe our selection process for choosing which \DNNs to include in our
\premodel.
The first \DNN we include is always the one which is optimal for most
of our training data.
We then look at the images which our current selection of \DNNs is unable to correctly classify, and add the \DNN which is most accurate
on these images.
We iteratively add \DNNs until our accuracy improvement is lower than a threshold $\theta$.
Using this method we are able to add \DNNs which best compliment one another when working together, maximizing the number of images
we can correctly classify with each new \DNN. Adding each \DNN in the order they appear in Figure~\ref{fig:optimum_models} would result
in a large overlap of correctly
classified images for each additional model. Below we will walk through the algorithm to show how we chose the model to include in our
\premodel.

We have chosen to set our threshold value, $\theta$ to 0.5, which is empirically decided during our pilot experiments.
 Figure~\ref{fig:optimum_models} shows the percentage of our training data
which considers each of our \CNNs to be optimal. There is a clear winner here, \mn{MobileNet\_v1\_100} is optimal for 70.75\% of our
training data, therefore it is chosen to be \textit{Model-1} for our \premodel. If we were to follow this convention and then choose the
next most optimal \CNN, we would choose \mn{Inception\_v1}. However, we do not do this as it would result in our \premodel being formulated
of many cheap, yet inaccurate models. Instead we choose to look at the training data on which our initial model (\textit{Model-1}) fails;
the remaining 29.25\% of our data.

\begin{figure*}[t!]
\def\arraystretch{0.82}
	\centering
	\begin{tabularx}{1\textwidth} {ccc}
		\includegraphics[width=0.32\textwidth]{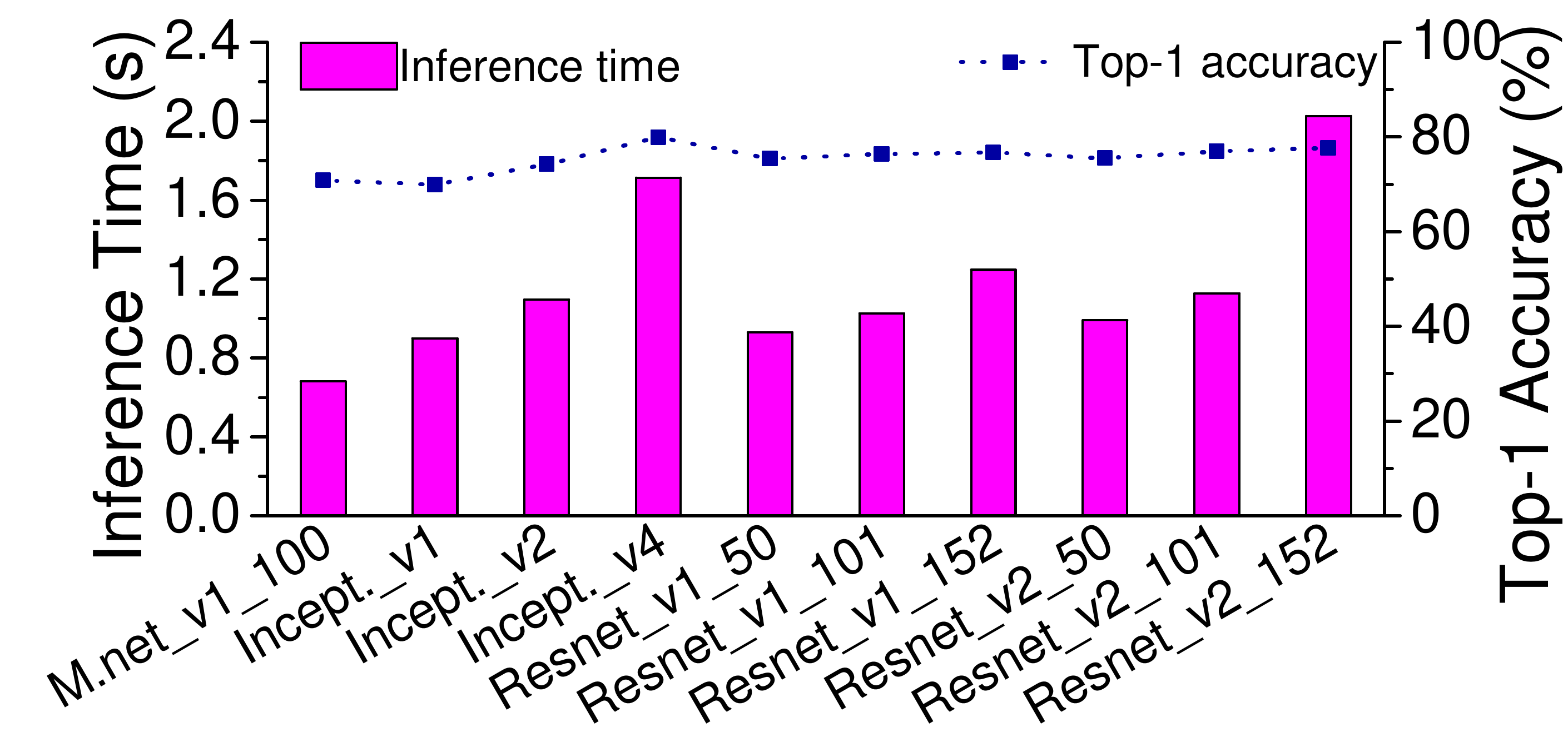} &
		\includegraphics[width=0.3\textwidth]{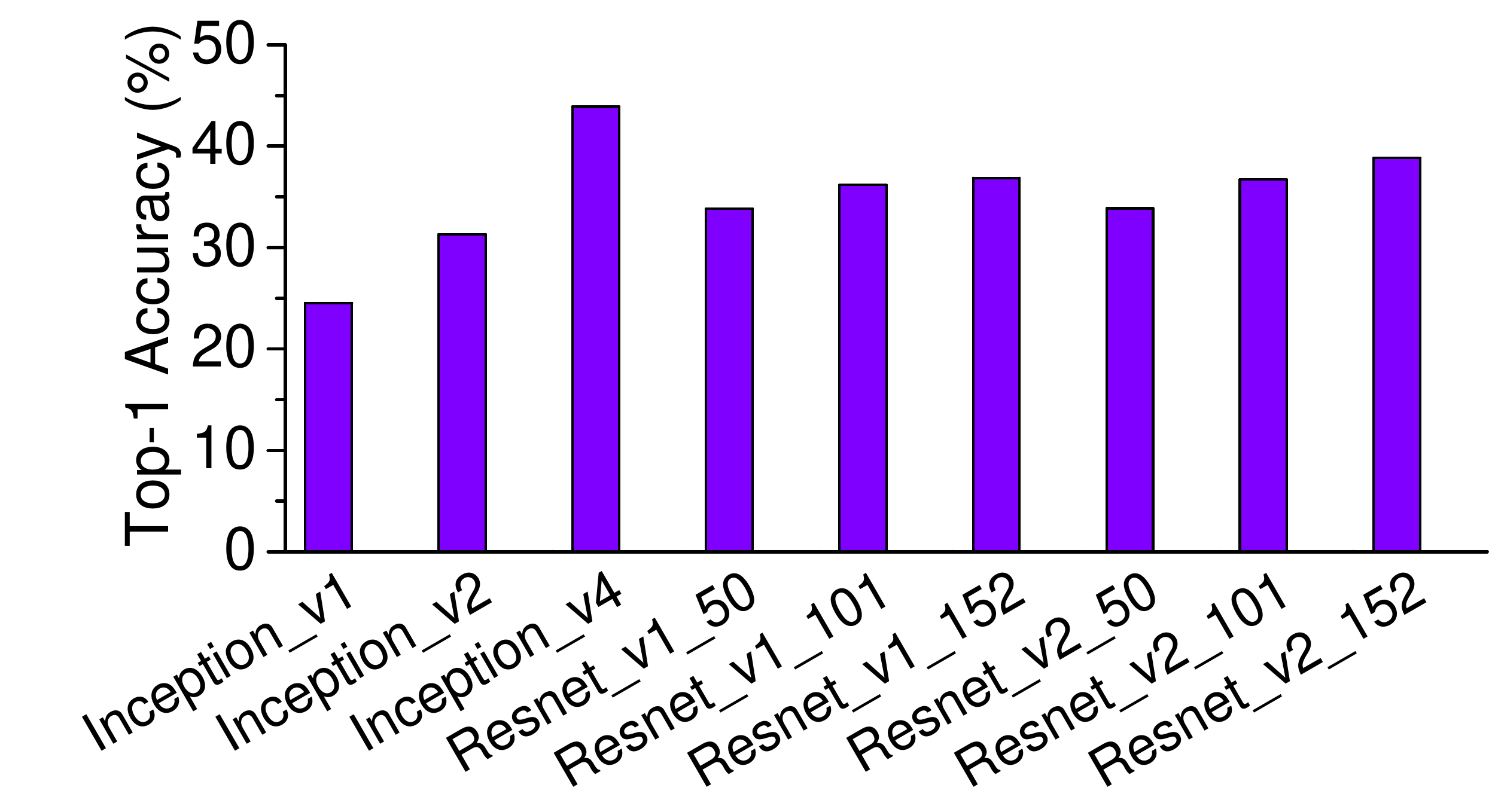}&
		\includegraphics[width=0.3\textwidth,clip]{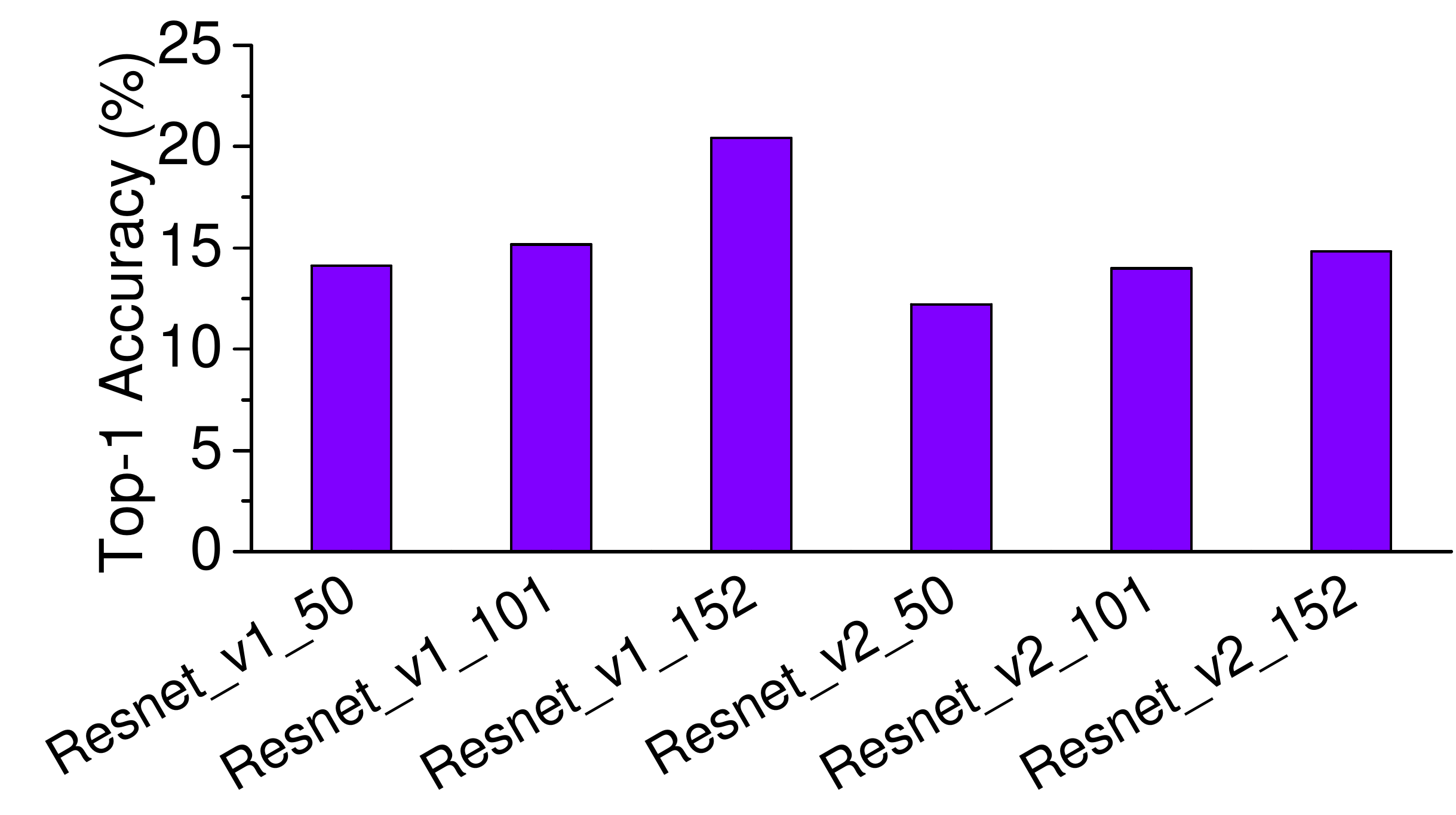}\\
		{\centering \scriptsize (a) \topone accuracy and inference time} &
		{\centering \scriptsize (b) \topone accuracy where \texttt{MobileNet} fails} &
		{\centering \scriptsize (c) \topone accuracy where \texttt{Mobilnet} \& \texttt{Inception} fails} \\

	\end{tabularx}
	\caption{(a) Shows the \topone accuracy and average inference time of all \CNNs considered in this work
	across our entire training dataset.
	(b) Shows the \topone accuracy of all \CNNs on the images on which \mn{MobileNet\_v1\_100} fails.
	(c) Shows the \topone accuracy of all \CNNs on the images on which \mn{MobileNet\_v1\_100} and \mn{Inception\_v4} fails.}
	\label{fig:classifier_selection}
\end{figure*}

We now exclusively consider the currently failing training data.
Figure~\ref{fig:classifier_selection}b shows the accuracy of our remaining \CNNs on the 29.25\% cases where \mn{MobileNet\_v1\_100}
fails. We can see that \mn{Inception\_v4} clearly wins here, correctly classifying 43.91\% of the remaining data; creating a 12.84\%
increase in \premodel accuracy; leaving 16.41\% of our data failing.
Repeating this process, shown in
Figure~\ref{fig:classifier_selection}c, we add \mn{ResNet\_v1\_152} to our \premodel, increasing total accuracy by
2.55\%. Finally we repeat this step one more time, to achieve a \premodel accuracy increase of <0.5, therefore <$\theta$, and terminate
here.

The result of this is a \premodel where: \textit{Model-1} is \mn{MobileNet\_v1\_100}, \textit{Model-2} is
\mn{Inception\_v4}, and, finally, \textit{Model-3} is \mn{ResNet\_v1\_152}.

\subsection{Training the \premodel} \label{sec:premodel_training}
Training our \premodel follows the standard procedure, and is a multi-step process. We describe the entire training process in detail
below, and provide a summary in Figure~\ref{fig:training}. Generally, we need to figure out which candidate \DNN is optimum for
each of our training images, we then train our model to predict the same for any \emph{new}, \emph{unseen} inputs.

\cparagraph{Generate Training Data.} Our training dataset consists of the feature values and the corresponding optimum
model for each image under an evaluation criterion. To evaluate the performance of the candidate \DNN models, they must be applied to
unseen images. We choose to use ILVRSC 2012 validation set, which contains 50k images, to generate training data for our \premodel.
This dataset provides a wide selection of images containing a range of topics and complexities.
We exhaustively execute each candidate model on the images, measuring the inference time and prediction results.
Inference time is measured on an unloaded machine to reduce noise; it is a one-off cost -- \ie it only needs to be completed once.
Because the relative runtime of models is stable, training can
be performed on a high-performance server to speedup data generation.

Using the execution time, \topone, and \topfive results we can calculate the \textit{optimum} classifier for each image; \ie
the model that achieves the accuracy goal (\topone or \topfive) in the least amount of time. Finally, we extract the feature values
(described in Section~\ref{sec:features}) from each image, and pair the feature values to the optimum classifier for each image,
resulting in our complete training dataset.

\cparagraph{Building the Model.} The training data is used to determine which classification models should be used and their optimal
hyper-parameters. Since we chose to use \NN  models to construct our \premodel, we use a standard supervized learning method
to train our \premodel. In \NN classification the training data is used to give a label to each point in the
model, then during prediction the model will use a distance measure (in our case we use Euclidian distance) to find the K nearest points
(in our case K=5). The label with the highest number of points to the prediction point is the output label.

\cparagraph{Training Cost.} Total training time of our \premodel is dominated by generating the  training data, which took less
than a day using a NVIDIA P40 GPU on a multi-core server. This can vary depending on the number of image classifiers to be
included. In our case, we had an unusually long training time as we considered 12 \DNN models. We would expect in deployment that the user
has a much smaller search space for image classifiers. The time in model selection and parameter tuning is negligible (less than 2 hours)
in comparison. See also Section~\ref{sec:overhead}.

\vspace{-1mm}
\subsection{Features} \label{sec:features}
One of the key aspects in building a successful predictor is developing the right features to characterize the input. In this work, we
considered a total of 30 candidate features, shown in Table~\ref{tbl:all_features}. The features were chosen based on previous image
classification work~\cite{hassaballah2016image} \eg edge based features (more edges lead to a more complex
 image), as well as intuition based on our motivation
(Section~\ref{sec:motivation}), \eg contrast (lower contrast makes it harder to see image content).

\begin{table}[t!]
\def\arraystretch{0.8}
	\centering
	\caption{Correlation values (absolute) of removed features to the kept ones.}
	\vspace{-2mm}
	\midsepremove
        \small
	\begin{tabular}{lll}
		\toprule
		\textbf{Kept Feature}												& \textbf{Removed Feature}            			&\textbf{Correl.}		\\
		\midrule
		\rowcolor{Gray}													& \textit{perceived\_brightness\_rms}				& 0.98 					\\
		\rowcolor{Gray}													& \textit{avg\_brightness}						& 0.91 					\\
		\rowcolor{Gray} 	\multirow{-3}{*}{\textit{avg\_perceived\_brightness}}	& \textit{brightness\_rms}						& 0.88		 			\\
							\textit{edge\_length1}				&      \textit{edge\_length \{4-7\}}						& 0.78 - 0.85 					\\						
		\rowcolor{Gray} 	\emph{hue1}												& \textit{hue \{2-6\}}								& 0.99 					\\
		\bottomrule
	\end{tabular}
	\label{tbl:feature_correlation}
\end{table}

\subsubsection{Feature selection} \label{sec:feature_sel}The time spent making a prediction is negligible in comparison to the overhead of feature extraction,
therefore by reducing our feature count we can decrease the total execution time of our \premodel. Moreover, by reducing the number of
features we are also improving the generalizability of our \premodel, \ie reducing the likelihood of over-fitting on our training
data.

Initially, we use correlation-based feature selection. If pairwise correlation is high for any pair of features, we drop one of them and keep the other; retaining most of the information.
We performed this by constructing a matrix of correlation coefficients using Pearson product-moment correlation (\textit{PCC}).
The coefficient value falls between $-1$ and $+1$.
The closer the absolute value is to $1$, the stronger the correlation between the two features being tested. We set a threshold of 0.75 and removed any features that had an absolute \textit{PCC} higher than the threshold.
Table~\ref{tbl:feature_correlation} summarizes the features we removed at this stage, leaving 17~features.

Next we evaluated the importance of each of our remaining features. To evaluate feature importance we first trained and evaluated our
\premodel using K-Fold cross validation (see also Section~\ref{sec:overhead}) and all of our current features, recording \premodel
accuracy. We then remove each feature and re-evaluate the model on the remaining features, taking note of the change in accuracy. If there
is a large drop in accuracy then the feature must be very important, otherwise, the features does not hold much importance for our
purposes. Using this information we performed a greedy search, removing the least important features one by one. By performing this search
we discovered that we can reduce our feature count down to 7 features (see Table~\ref{tbl:chosen_features}) while having very little impact
on our model accuracy. Removing any of the remaining 7 features resulted in a significant drop in model accuracy.

\begin{table}[t!]
\def\arraystretch{0.82}
	\small
	\centering
    	\caption{The chosen features.}
	\vspace{-3mm}
	\midsepremove
	\begin{tabular}{lll}
		\toprule
		\rowcolor{Gray}		\textit{n\_keypoints}           & \textit{avg\_perceived\_brightness}	& \textit{hue1}              \\
							\textit{contrast}		        & \textit{area\_by\_perim}				& \textit{edge\_length1}     \\
		\rowcolor{Gray}		\textit{aspect\_ratio} 	        & 										&                            \\
		\bottomrule
	\end{tabular}
	\label{tbl:chosen_features}
\end{table}

\subsubsection{Feature scaling} The final step before passing our features to a machine learning model is scaling each of the features to a common range (between 0 and 1) in order to prevent the range of any single feature being a factor in its importance.
Scaling features does not affect the distribution or variance of their values.
To scale the features of a new image during deployment we record the minimum and maximum values of each feature in the training dataset, and use these to scale the corresponding features.

\begin{figure}[t!]
	\centering
	\includegraphics[width=0.7\textwidth]{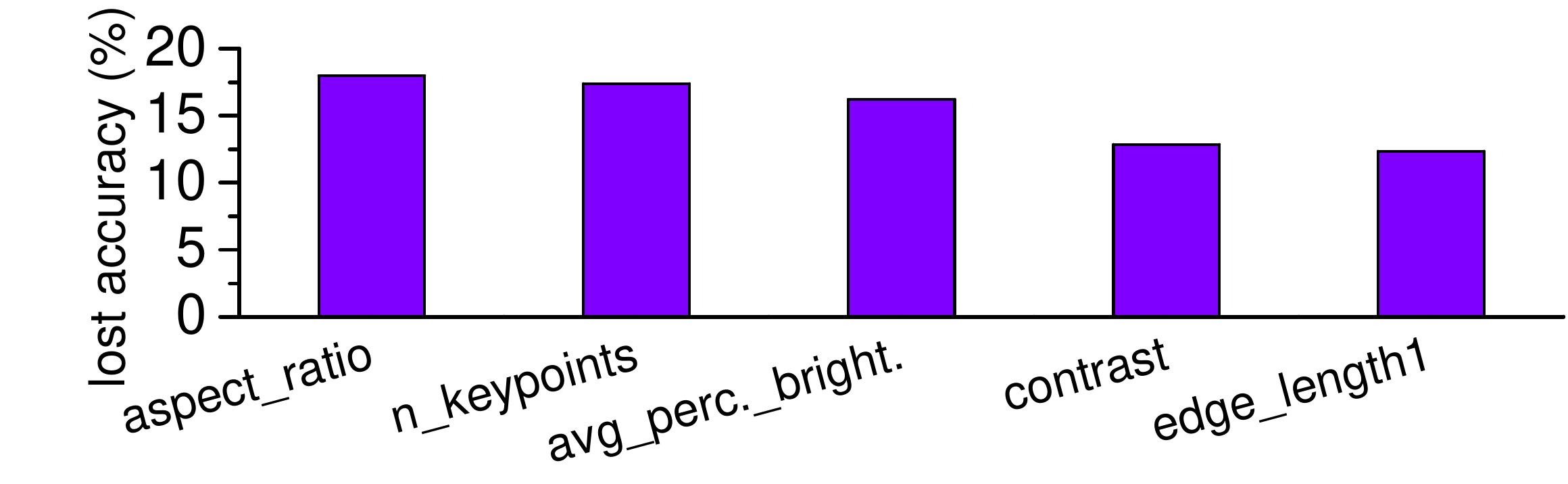}\\
    \vspace{-2mm}
	\caption{The top five features which can lead to a high loss in accuracy if they are not used in our \premodel.}
    \vspace{-2mm}
	\label{fig:top_5_feature_importance}
\end{figure}

\subsubsection{Feature analysis}
Figure~\ref{fig:top_5_feature_importance} shows the top 5 dominant features based on their impact on our \premodel accuracy.
We calculate feature importance by first training a \premodel using all 7 of our chosen features, and note the accuracy of our model.
In turn, we then remove each of our features, retraining and evaluating our \premodel on the other 6, noting the drop in accuracy.
We then normalize the values to produce a percentage of importance for each of our features.
It is clear our features hold a very similar level of importance, ranging between 18\% and 11\% for our most and least important feature respectively.
The similarity of feature importance is an indication that each of our features is able to represent distinct information about each image.
all of which is important for the prediction task at hand.

\subsection{Runtime Deployment}
Deployment of our proposed method is designed to be simple and easy to use, similar to current image
classification techniques.
We have encapsulated all of the inner workings, such as needing to read the output of the \premodel and then
choosing the correct image classifier.
A user would interact with our proposed method in the same way as any other image classifier:
simply calling a prediction function and getting the result in return as predicted labels and their confidence levels.

\section{Experimental Setup}
\subsection{Platform and Models}
\cparagraph{Hardware.} We evaluate our approach on the NVIDIA Jetson TX2 embedded deep learning platform. The system has a 64~bit dual-core
Denver2  and a 64~bit quad-core ARM Cortex-A57 running at 2.0~Ghz, 
and a 256-core NVIDIA Pascal GPU running at 1.3~Ghz. The board has 8~GB
of LPDDR4 RAM and 96~GB of storage (32~GB eMMC plus 64~GB SD card).

\cparagraph{System Software.} Our evaluation platform runs Ubuntu 16.04 with Linux kernel v4.4.15. We use Tensorflow v.1.0.1, cuDNN (v6.0)
and CUDA (v8.0.64). Our \premodel is implemented using the Python scikit-learn package. Our feature extractor is built upon OpenCV and
SimpleCV.

\cparagraph{Deep Learning Models.} We consider 14 pre-trained \CNN models for image recognition from the TensorFlow-Slim
library~\cite{silberman2013tensorflow}. The models are built using TensorFlow and trained on the ImageNet ILSVRC 2012 training set.

\subsection{Evaluation Methodology \label{sec:method}}

\cparagraph{Model Evaluation.} We use \emph{10-fold cross-validation} to evaluate our \premodel on the ImageNet ILSVRC 2012 validation set.
Specifically, we partition the 50K validation images into 10 equal sets, each containing 5K images. We retain one set for testing our
\premodel, and the remaining 9 sets are used as training data. We repeat this process 10 times (folds), with each of the 10 sets used
exactly once as the testing data. This standard methodology evaluates the generalization ability of a machine-learning model.

We evaluate our approach using the following metrics:

\begin{itemize}
\item \emph{\textbf{Inference time} (lower is better)}. Wall clock time between a model taking in an input and producing an output,
    including the overhead of our \premodel.

\item \emph{\textbf{Energy consumption} (lower is better)}. The energy used by a model for inference. For our approach, this also
    includes the energy consumption of the \premodel. We deduct the static power used by the hardware when the system is idle.

\item \emph{\textbf{Accuracy} (higher is better)}. The ratio of correctly labeled images to the total number of testing images.

\item \emph{\textbf{Precision} (higher is better)}. The ratio of a correctly predicted images to the total number of images that are predicted to have a
    specific object. This metric answers e.g., ``\emph{Of all the images that are labeled to have a cat, how many actually have a cat?}".

\item \emph{\textbf{Recall} (higher is better)}. The ratio of correctly predicted images to the total number of test images that belong to an object class.
    This metric answers e.g., ``\emph{Of all the test images that have a cat, how many are actually labeled to have a cat?}".

\item \emph{\textbf{F1 score} (higher is better)}.  The weighted average of Precision and Recall, calculated as $2\times\frac{Recall
    \times Precision} {Recall + Precision}$. It is useful when the test datasets have an uneven distribution of object classes.

\end{itemize}

\cparagraph{Performance Report.} We report the \emph{geometric mean} of the aforementioned evaluation metrics across the cross-validation
folds. To collect inference time and energy consumption, we run each model on each input repeatedly until the 95\% confidence bound per
model per input is smaller than 5\%. In the experiments, we exclude the loading time of the \CNN models as they only need to be loaded once
in practice. However, we include the overhead of our \premodel in all our experimental data. To measure energy consumption, we developed a
lightweight runtime to take readings from the on-board energy sensors at a frequency of 1,000 samples per second.
It is to note that our work does not directly optimise for energy consumption. We found that in our scenario there is  little difference
when optimizing for energy consumption compared to time.

\section{Experimental Results \label{sec:results}}
\begin{figure*}[t!]
\def\arraystretch{0.8}
	\centering
	\begin{tabularx}{1\textwidth} {>{\centering\arraybackslash}m{1.6in}>{\centering\arraybackslash}m{1.6in}>{\centering\arraybackslash}m{1.6in}>{\centering\arraybackslash}m{1.6in}}

		\includegraphics[width=0.249\textwidth,clip]{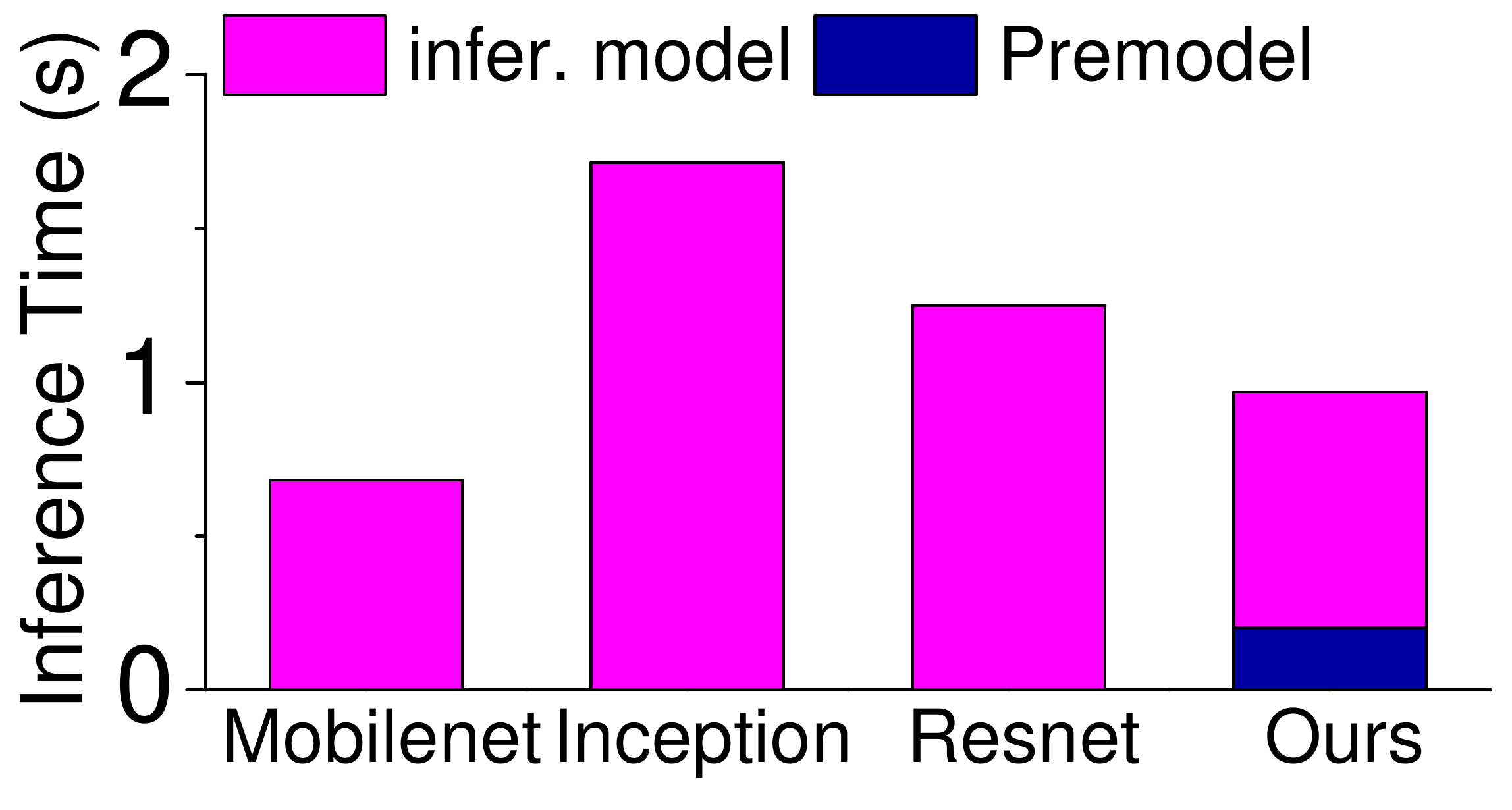} &
		\includegraphics[width=0.249\textwidth,clip]{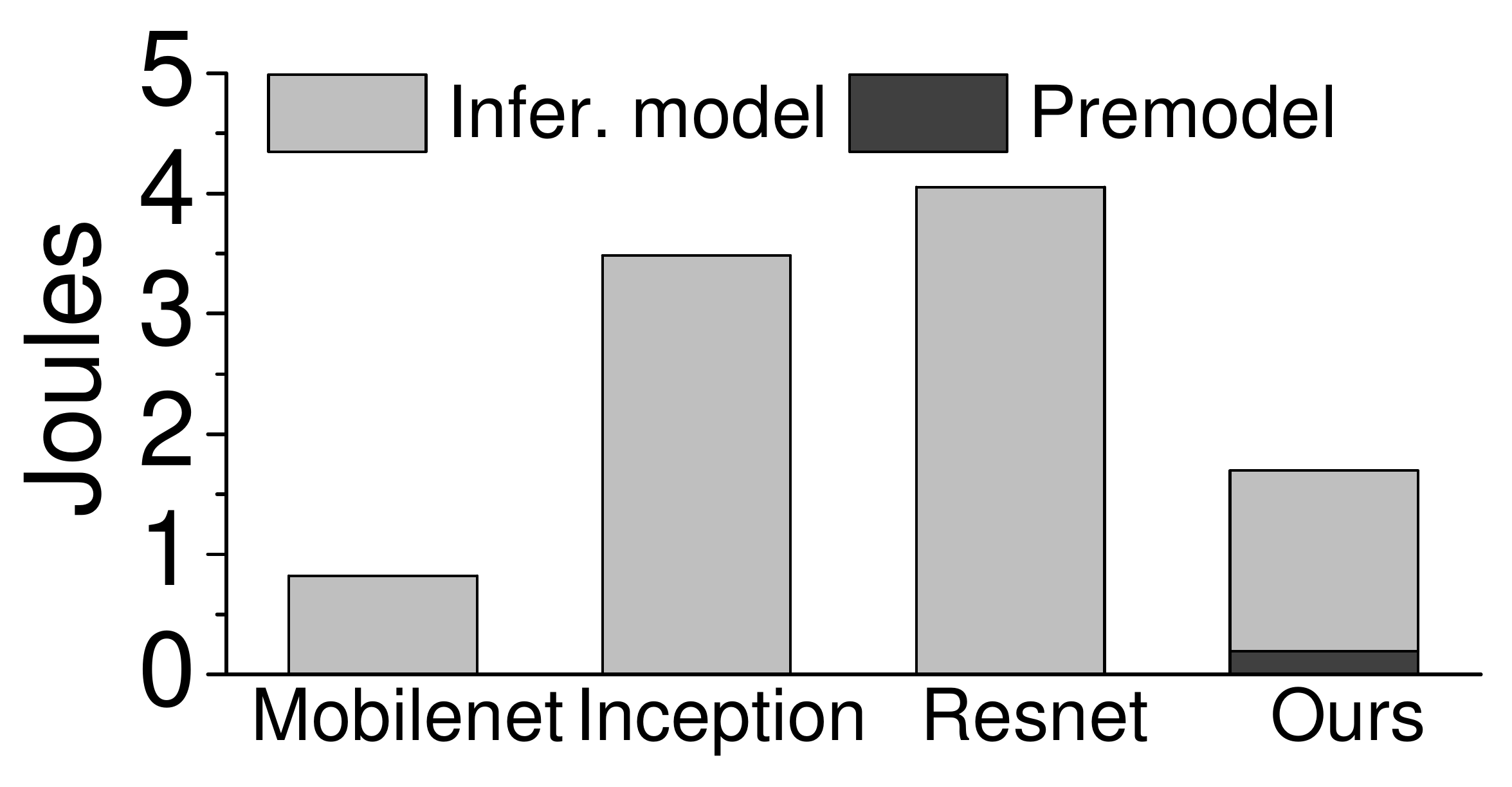} &	
		\includegraphics[width=0.249\textwidth,clip]{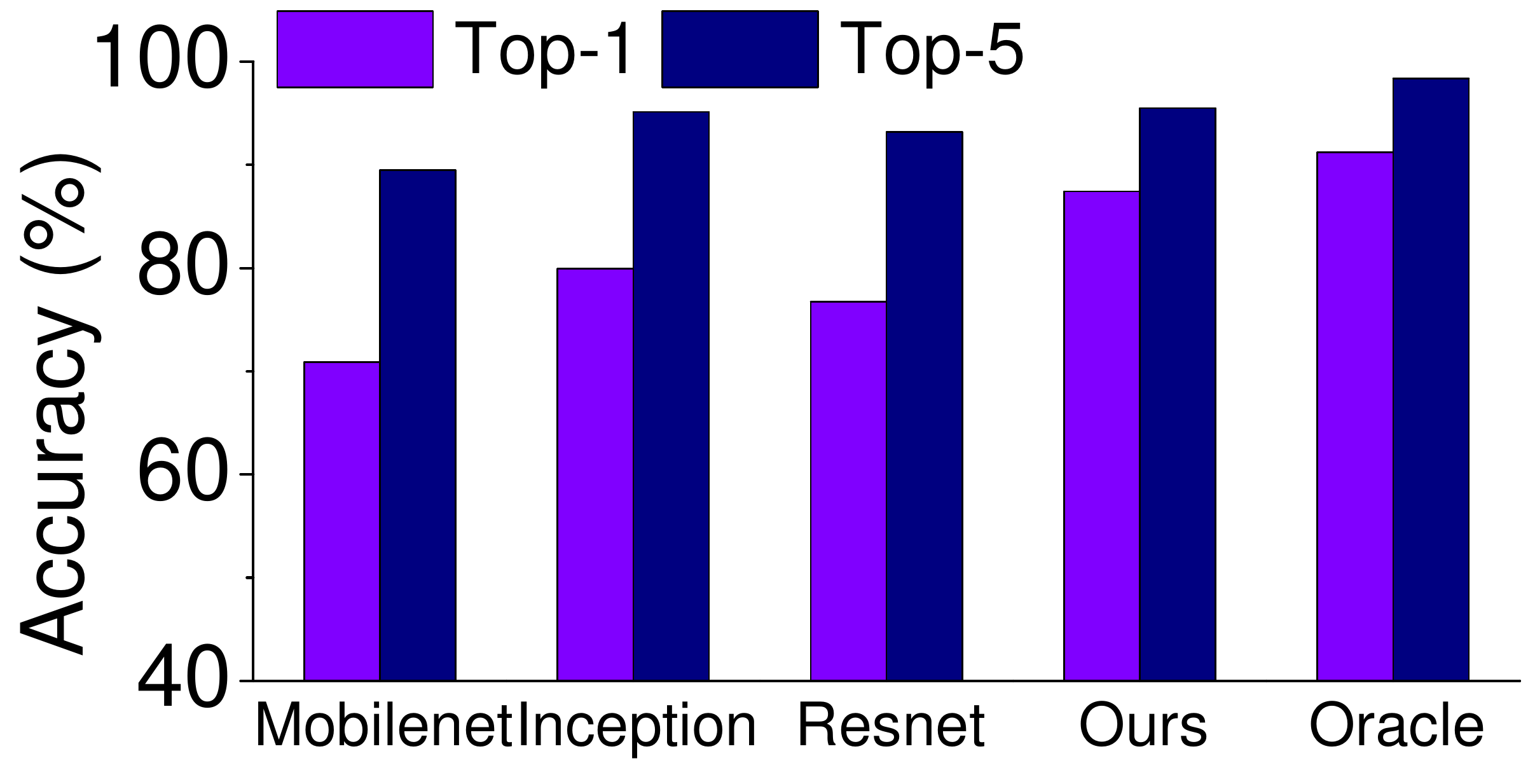} &
		\includegraphics[width=0.249\textwidth,clip]{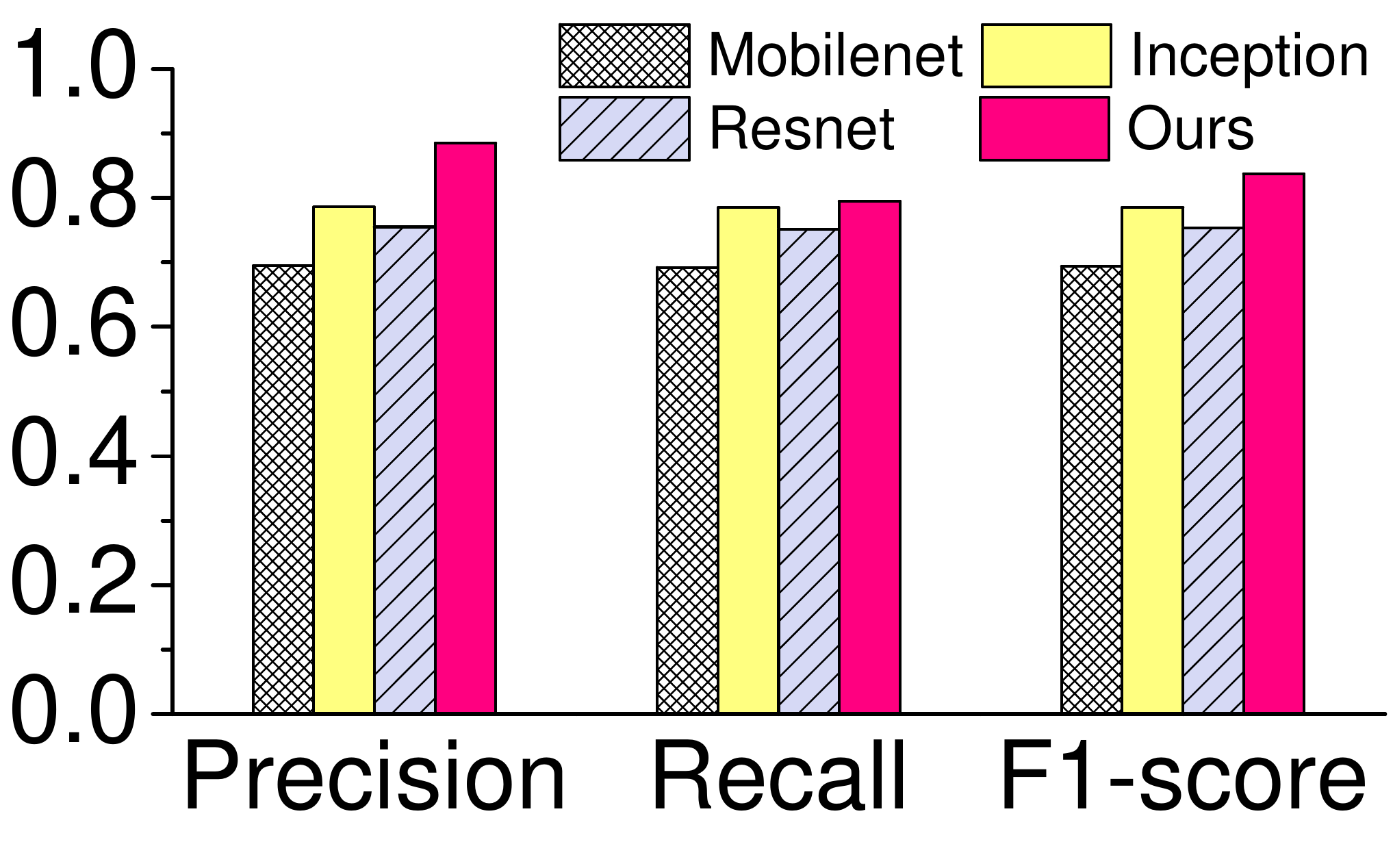} \\
		{\centering \scriptsize (a) Inference Time} &
		{\centering \scriptsize (b) Energy Consumption } &
		{\centering \scriptsize (c) Accuracy} 	&	{\centering \scriptsize (d) Precision, Recall \& F1 score} \\
		
	\end{tabularx}
	\caption{
    Overall performance of our approach against individual models for inference time (a), energy consumption (b), accuracy (c), precision,
    recall and F1 score (d). Our approach gives the best overall performance.
	 }
	\label{fig:expermiental_results}
\end{figure*}

\subsection{Overall Performance}
\cparagraph{Inference Time.} Figure~\ref{fig:expermiental_results}a compares the inference time among individual \DNN models and our
approach. \texttt{MobileNet} is the fastest model for inferencing, being 2.8x and 2x faster than \texttt{Inception} and \texttt{ResNet},
respectively, but is least accurate (see Figure~\ref{fig:expermiental_results}c). Our \premodel alone is 3x faster than \texttt{MobileNet}.
Most the overhead of our \premodel comes from feature extraction. The average inference time of our approach is under a second, which is
slightly longer than the 0.7 second average time of \texttt{MobileNet}. Our approach is 1.8x faster than \texttt{Inception}, the most
accurate inference model in our model set. Given that our approach can significantly improve the prediction accuracy of \texttt{Mobilenet},
we believe the modest cost of our \premodel is acceptable.

\cparagraph{Energy Consumption.} Figure~\ref{fig:expermiental_results}b gives the energy consumption. On the Jetson TX2 platform, the
energy consumption is proportional to the model inference time. As we speed up the overall inference, we reduce the energy consumption by
more than 2x compared to \texttt{Inception} and \texttt{Resnet}. The energy footprint of our \premodel is small, being 4x and 24x lower
than \texttt{MobileNet} and \texttt{ResNet} respectively. As such, it is suitable for power-constrained devices, and can be used to
improve the overall accuracy when using multiple inferencing models. Furthermore, in cases where the \premodel predicts that none of the
\DNN models can successfully infers an input, it can skip inference to avoid wasting power. It is to note that since our \premodel runs on
the CPU, its energy footprint ratio is smaller than that for runtime.

\cparagraph{Accuracy.} Figure~\ref{fig:expermiental_results}c compares the \topone and \topfive accuracy achieved by each approach. We
also show the best possible accuracy given by a \emph{theoretically} perfect predictor for model selection, for which we call
\texttt{Oracle}. Note that the \texttt{Oracle} does not give a 100\% accuracy because there are cases where all the \DNNs fail.
However, not all \DNNs fail on the same images, \ie \texttt{ResNet} will successfully classify some images which
\texttt{Inception} will fail on.
Therefore, byeffectively leveraging multiple models, our approach outperforms all individual inference models.

It improves the accuracy of
\texttt{MobileNet} by 16.6\% and 6\% respectively for the \topone and the \topfive scores. It also improves the \topone accuracy of
\texttt{ResNet} and \texttt{Inception} by 10.7\% and 7.6\% respectively. While we observe little improvement for the \topfive score over
\texttt{Inception} -- just 0.34\% -- our approach is 2x faster than it. Our approach delivers over 96\% of the \texttt{Oracle} performance
(87.4\% vs 91.2\% for \topone and 95.4\% vs 98.3\% for \topfive). Moreover, our approach never picks a model that fails while others can success. This
result shows that our approach can improve the inference accuracy of individual models.

\cparagraph{Precision, Recall, F1 Score.} Finally, Figure~\ref{fig:expermiental_results}d shows our approach outperforms individual \DNN
models in other evaluation metrics. Specifically, our approach gives the highest overall precision, which in turns leads to the best F1
score. High precision can reduce false positive, which is important for certain domains like video surveillance because it can reduce the
human involvement for inspecting false positive predictions.

\subsection{Alternative Techniques for Premodel} \label{sec:alt_premodels}
\begin{figure}[t!]
	\centering
	\includegraphics[width=0.7\textwidth]{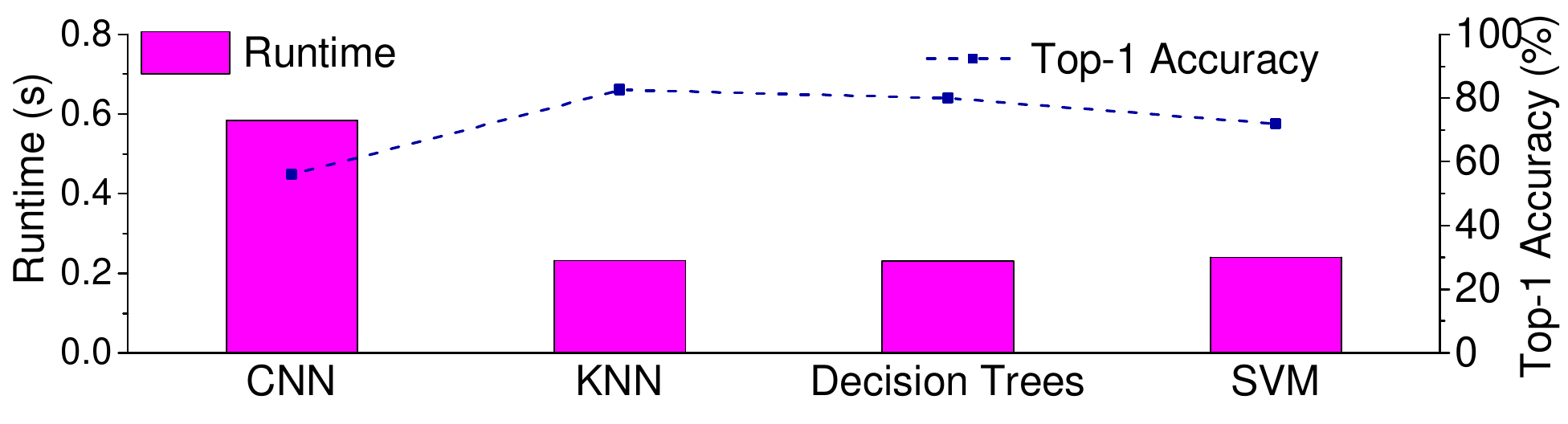}
	\caption{Comparison of alternative predictive modeling techniques for building the \premodel.}
	\label{fig:alternative_techniques}
\end{figure}

Figure~\ref{fig:alternative_techniques} shows the \topone accuracy and runtime for using different techniques to construct the \premodel.
Here, the learning task is to predict which of the inference models, \texttt{MobileNet}, \texttt{Inception}, and \texttt{ResNet}, to use.
In addition to \NN, we also consider \CNNs, Decision Trees (\DT) and Support Vector Machines (\SVM). We use the MobileNet structure, which
is designed for embedded inference, to build the \CNN-based \premodel. We train all the models using the same training examples. We also
use the same feature set for the \NN, \DT, and \SVM. For the \CNN, we use a hyperparameter tuner \cite{klein2016fast} to optimize the
training parameters, and we train the model for over 500 epochs.

While we hypothesized a \CNN model to be effectively in predicting from an image to the output, the results are disappointing given its
high runtime overhead. Our chosen \NN model has an overhead that is comparable to the \DT and the \SVM, but has a higher accuracy. It is
possible that the best technique can change as the application domain and training data size changes, but our generic approach for feature
selection and model selection remains applicable.

\begin{figure}[t!]
	\centering
	\includegraphics[width=0.7\textwidth]{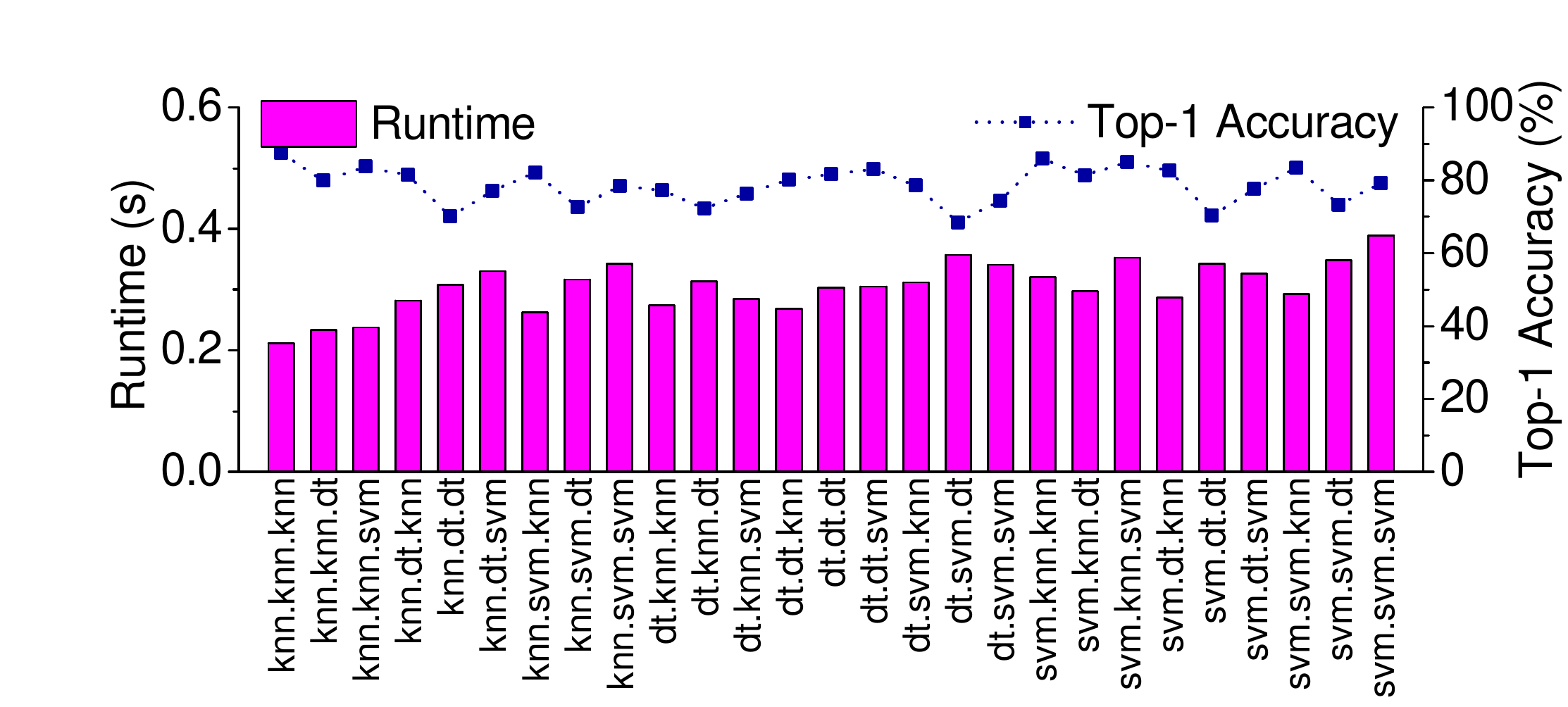}
	\caption{Using different modeling techniques to form a 3-level \premodel.}
    \vspace{-2mm}
    \label{fig:pre_model_comparison}
\end{figure}

Figure~\ref{fig:pre_model_comparison} shows the runtime and \topone accuracy by using the \NN, \DT and \SVM to construct a three level hierarchical \premodel configuration denoted as $X.Y.Z$, where $X$, $Y$ and $Z$ indicate the modeling technique for the first,
second and third level of the \premodel, respectively. The result shows that our chosen \premodel organization,  (\ie, \NN.\NN.\NN), has
the highest \topone accuracy (87.4\%) and the fastest running time (0.20~second). One of the benefits of using a \NN model in all levels is
that the neighboring measurement only needs to be performed once as the results can be shared among models in different levels; \ie
the runtime overhead is nearly constant if we use the \NN across all hierarchical levels.
The accuracy for each of our \NN models in our \premodel is 95.8\%, 80.1\%, 72.3\%, respectively.

\begin{figure}
	\includegraphics[width=0.7\textwidth]{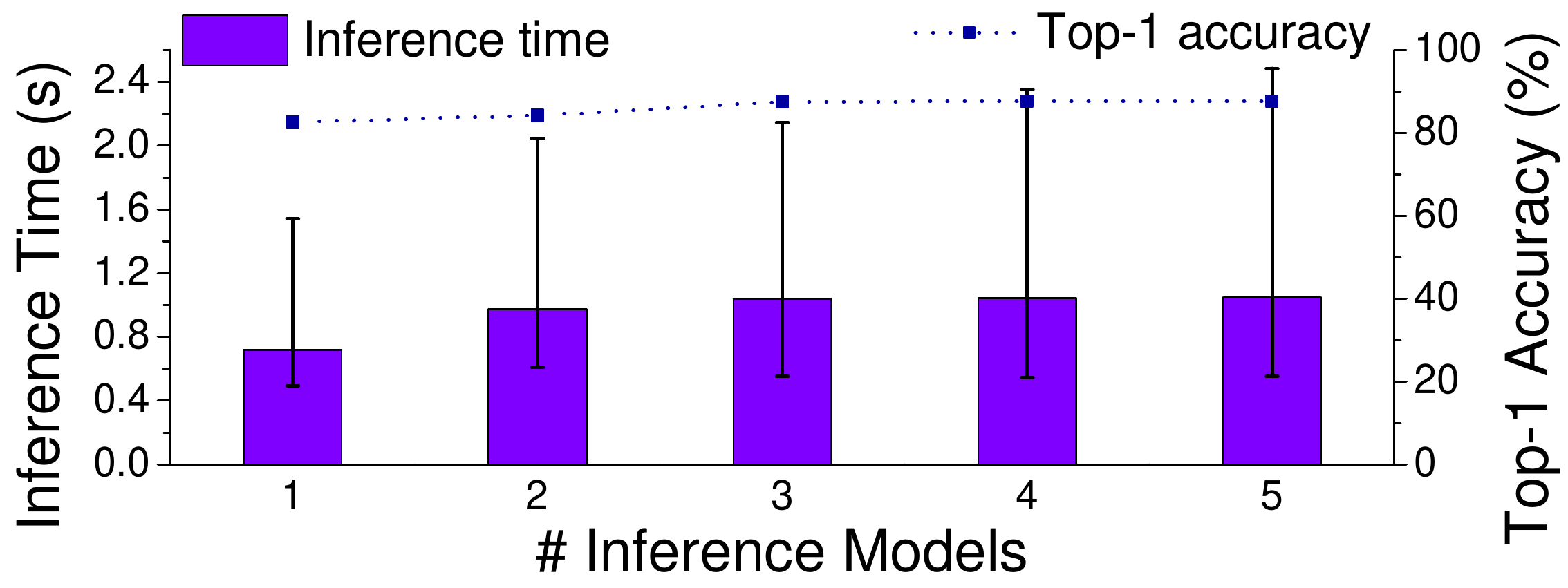}
	\caption{Overhead and achieved performance when using different numbers of \DNN models. The min-max bars show the range of
    inference time across testing images.}
	\label{fig:pre_model_different_levels}
\end{figure}

\subsection{Impact of Inference Model Sizes}
In Section~\ref{sec:classifier_selection} we describe the method we use to chose which \DNN models to include. Using this method, and
temporarily ignoring the model selection threshold $\theta$ in Algorithm~\ref{alg:classifier_selection}, we constructed
Figure~\ref{fig:pre_model_different_levels}, where we compare the \topone accuracy and execution time using up to 5 \NN models. As we
increase the number of inference models, there is an increase in the end to end inference time as expensive models are more likely to be
chosen. At the same time, however, the \topone accuracy reaches a plateau of ($\approx$87.5\%) by using three \NN models. We conclude that
choosing three \NN models would be the optimal solution for our case, as we are no longer gaining accuracy to justify the increased cost.
This is in line with our choice of a value of 0.5 for $\theta$.

\begin{figure}[t!]
	\centering
	\includegraphics[width=0.7\textwidth]{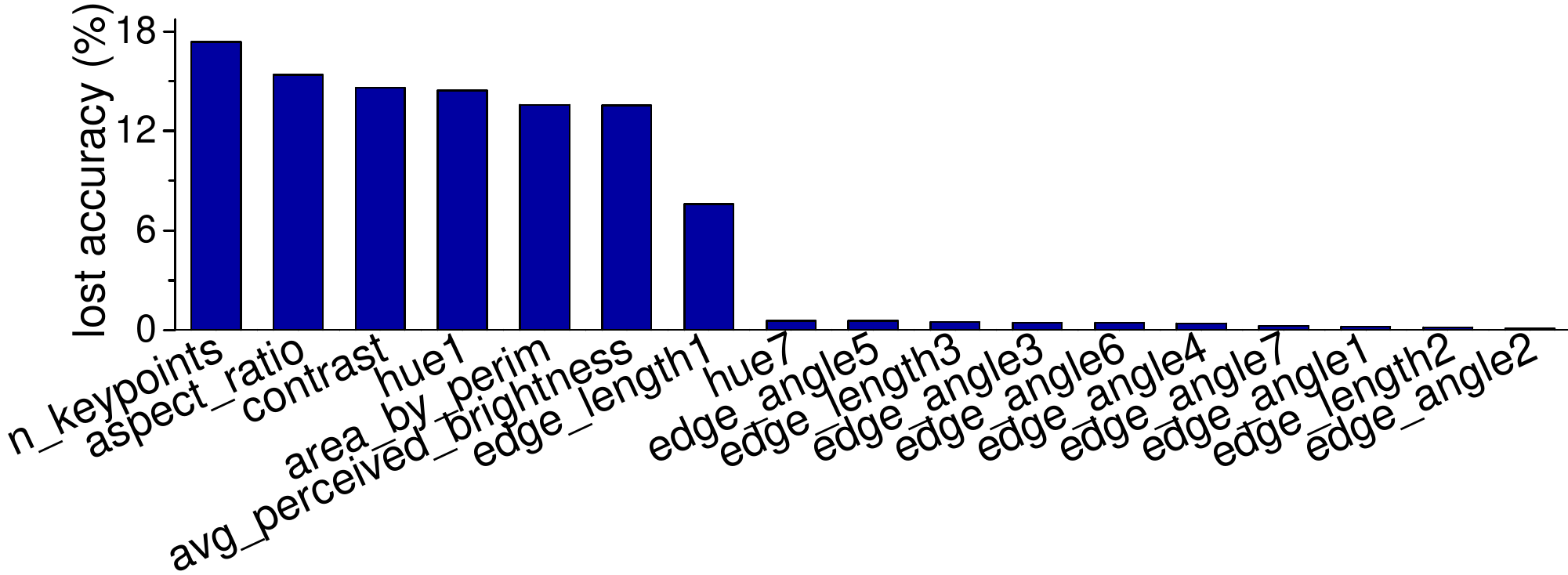}\\
	\caption{Accuracy loss if a feature is not used.}
    \vspace{-2mm}
	\label{fig:feature_17_importance}
\end{figure}

\begin{figure}[t!]
	\centering
	\includegraphics[width=0.7\textwidth]{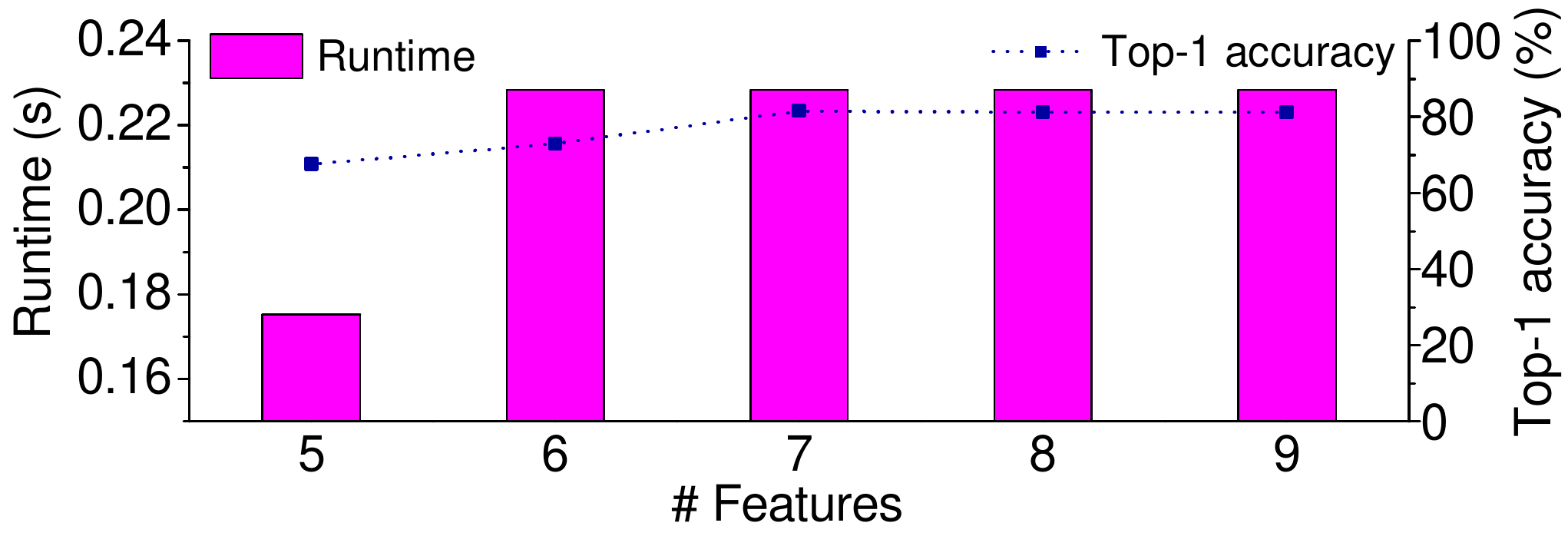}\\
    \vspace{-2mm}
	\caption{Impact of feature sizes.}
    \vspace{-2mm}
	\label{fig:feature_removal}
\end{figure}

\subsection{Feature Importance}

In Section~\ref{sec:features} we describe our feature selection process, which resulted in using 7 features to represent each image to our
\premodel. In Figure~\ref{fig:feature_17_importance} we show the importance of all of our considered features which were not removed by our
correllation check, shown in Table~\ref{tbl:feature_correlation}. Upon observation it is clear that the 7 features we have chosen to keep
are the most important; there is a sudden drop in feature importance at feature 8 (\textit{hue7}). Furthermore, in
Figure~\ref{fig:feature_removal} we show the impact on \premodel execution time and \topone accuracy when we change the number of features
we use. By decreasing the number of features there is a dramatic decrease \topone accuracy, with very little change in extraction time. To
reduce overhead, we would need to reduce our feature count to 5, however this comes at the cost of a 13.9\% decrease in \topone accuracy.
By increasing the feature count it can be seen that there is minor changes in overhead, but, surprisingly, there is actually also a small
decrease in \topone accuracy of 0.4\%. From this we can conclude that using 7 features is ideal.

\subsection{Training and Deployment Overhead \label{sec:overhead}}
Training the \premodel is a \emph{one-off} cost, and is dominated by the generation of training data which takes in total less than a day
(see Section~\ref{sec:premodel_training}). We can speed this up by using multiple machines. However, compared to the training time
of a typical \DNN model, our training overhead is negligible.

The runtime overhead of our \premodel is minimal, as depicted in Figures~\ref{fig:expermiental_results}a. Out of a total average execution
time of less than a second to classify an image, our \premodel accounts for 20\%. Compared to the most (\mn{ResNet\_v2\_152}) and least
(\mn{MobileNet}) expensive models, this translates to 9.52\% and 27\%, respectively. Furthermore, our energy footprint is smaller, making
up 11\% of the total cost. Comparing this to the most and least expensive models, again, gives an overhead of 7\% and 25\%, respectively.
Finally, while we primarily consider inference time and accuracy, we acknowledge that RAM may be limited on an embedded system. We found,
however, that our chosen \DNN models only use 25\% of the available RAM on our platform. The memory footprint of our \premodel is
negligible.

\section{Discussion}
Naturally there is  room for further work and possible improvements. We discuss a few points here.

\cparagraph{Alternative Domains.} This work focuses on \CNNs because it is a commonly used deep learning architecture. To extend our work
to other domains and recurrent neural networks (\RNN), we would need a new set of features to characterize the input, e.g., text embeddings
for machine translation~\cite{zou2013bilingual}. However, our automatic approach on feature selection and \premodel construction remains
applicable.

%
\cparagraph{Feature Extraction. } The majority of our overhead is caused by feature extraction
 for our \premodel. Our prototype feature extractor is written in Python; by re-writing this tool in a more
efficient language can reduce the overhead. There are also hotshots in our code which would benefit from parallelism.

\cparagraph{Soundness.} It is possible that our \premodel will provide an incorrect prediction. That is, it could choose either a \DNN that
gives an incorrect result, or a more expensive \DNN. However, by using the feature distance as a confidence measurement, we can have a
degree of soundness guarantee.


\cparagraph{Processor Choice. } By default, inference is carried out on a GPU, but this may not always be the best choice.  Previous work
has already shown machine learning techniques to be successful at selecting the optimal computing device~\cite{taylor2017adaptive}. This
can be integrated into our existing learning framework.

\cparagraph{Model Size.} Our approach uses multiple pre-trained \DNN models for inference. In comparison to the default method of simply
using a single model, our approach would require more storage space. A solution for this would involve using model compression techniques
to generate multiple compressed models from a single accurate model. Each compressed model would be smaller and is specialized at certain
tasks. The result of this is numerous models share many weights in common, which allows us to allowing us to amortize the cost of using
multiple models.

\section{Related Work}
\DNNs have shown astounding successes in various tasks that previously seemed
difficult~\cite{alexnet,Lee:2009:UFL:2984093.2984217,cho2014learning}. Despite the fact that many embedded devices require precise sensing
capabilities, adoption of \DNN models on such systems has notably slow progress. This mainly due to \DNN-based
inference being typically a computation intensive task, which inherently runs slowly on embedded devices due to limited resources.

Methods have been proposed to reduce the computational demands of a deep model by trading prediction accuracy for runtime, compressing a
pre-trained network~\cite{DBLP:journals/corr/JinDC14,han2015learning,Chen:2015:CNN:3045118.3045361,
springenberg2014striving,DBLP:journals/corr/IandolaMAHDK16,DBLP:journals/corr/RastegariORF16}, training small networks
directly~\cite{projectnet,Georgiev:2017:LMA:3139486.3131895}, or a combination of both~\cite{howard2017mobilenets}. Using these approaches,
a user now needs to decide when to use a specific model, in order to meet the prediction accuracy requirement with minimal latency.
It is a non-trivial task to make such a crucial decision as the application context (e.g. the model input) is often unpredictable and
constantly evolving. Our work alleviates the user burden by automatically selecting an appropriate model to use.

Neurosurgeon \cite{Kang2017neurosurgeon} identifies when it is beneficial (\eg in terms of energy consumption and end-to-end latency) to offload a \DNN layer to be computed on the cloud. Unlike
Neurosurgeon, we aim to minimize \emph{on-device} inference time without compromising prediction accuracy.
The Pervasive \CNN~\cite{7920809} generates multiple computation kernels for each layer of a \CNN, which are then dynamically selected
according to the inputs and user constraints. A similar approach presented in \cite{RodriguezWZMH17} trains a model twice, once on shared
data and again on personal data, in an attempt to prevent personal data being sent outside the personal domain. In contrast to the latter
two works, our approach allows having a diverse set of networks, by choosing the most effective network to use at runtime. They, however,
are complementary to our approach, by providing the capability to fine-tune a single network structure.

Recently, a number of software-based approaches have been proposed to accelerate \CNNs on embeded devices. They aim to accelerate inference
time by exploiting parameter tuning~\cite{latifi2016cnndroid}, computational kernel optimization~\cite{han2016eie,
bhattacharya2016sparsification}, task parallelism~\cite{Motamedi:2017:MIR:3145508.3126555,lane2016deepx,rallapalli2016very}, and trading
precision for time~\cite{Huynh:2017:DMG:3081333.3081360} etc. Since a single model is unlikely to meet all the constraints of accuracy,
inference time and energy consumption across inputs~\cite{CanzianiPC16,guo2017towards}, it is attractive to have a strategy to dynamically
select the appropriate model to use. Our work provides exactly such a capability and is thus complementary to these prior approaches.

Off-loading computation to the cloud can accelerate \DNN model inference \cite{teerapittayanon2017distributed}, but this is not always
applicable due to privacy, latency or connectivity issues. The work presented by Ossia \etal partially addresses the issue of
privacy-preserving when offloading \DNN inference to the cloud ~\cite{ossia2017hybrid}. Our adaptive model selection approach allows one to
select which model to use based on the input, and is also useful when cloud offloading is prohibitively because of the latency requirement
or the lack of connectivity.

Machine learning has been employed for various optimization tasks~\cite{mlcpieee,Samreen2016Daleel}, including code
optimization~\cite{wang2014integrating,Tournavitis:2009:THA:1542476.1542496,Wang:2009:MPM:1504176.1504189,wang2010partitioning,grewe2013portable,wang2013using,DBLP:journals/taco/WangGO14,taylor2017adaptive,
ogilvie2014fast,cummins2017end,ogilvie2017minimizing,ipdpsz18,spmv}, task
scheduling~\cite{grewe2011workload,emani2013smart,grewe2013opencl,Delimitrou:2014:QRQ:2541940.2541941,ren2017optimise}, etc.  Our approach
is closely related to ensemble learning where multiple models are used to solve an optimization problem. This technique is shown to be
useful on scheduling parallel tasks~\cite{Emani:2015:CDM:2737924.2737999} and optimize application memory
usage~\cite{Marco:2017:ISA:3135974.3135984}. This work is the first attempt in applying this technique to optimize deep inference on
embedded devices.

\section{Conclusion}
This paper has presented a novel scheme to dynamically select a deep learning model to use on an embedded device. Our approach provides a
significant improvement over individual deep learning models in terms of accuracy, inference time, and energy consumption. Central to our
approach is a machine learning based method for deep learning model selection based on the model input and the precision requirement. The
prediction is based on a set of features of the input, which are tuned and selected by our automatic approach. We apply our approach to the
image recognition task and evaluate it on the Jetson TX2 embedded deep learning platform using the ImageNet ILSVRC 2012 validation dataset.
Experimental results show that our approach achieves an overall \topone accuracy of above 87.44\%, which translates into an improvement of
 7.52\% and 1.8x reduction in inference time when compared to the most-accurate single deep learning model.

\balance
\bibliographystyle{ACM-Reference-Format}
\bibliography{bibs,zheng}

\balance

\end{document}